\documentclass[journal]{IEEEtran}

\ifCLASSINFOpdf
\else
   \usepackage[dvips]{graphicx}
\fi
\usepackage{url}

\usepackage{graphicx}
\usepackage{booktabs}
\usepackage{subcaption}
\usepackage{bm}
\usepackage{tikz}
\usepackage{adjustbox}
\newcommand{\ra}[1]{\renewcommand{\arraystretch}{#1}}

\newcommand*{\tikzbullet}[2]{%
  \setbox0=\hbox{\strut}%
  \begin{tikzpicture}
    \useasboundingbox (-.25em,0) rectangle (.25em,\ht0);
    \filldraw[draw=#1,fill=#2] (0,0.5\ht0) circle[radius=.25em];
  \end{tikzpicture}%
}

\usepackage{amsmath}
\usepackage{amssymb}    
\newcommand{\sign}{\text{sign}}

\usepackage{dblfloatfix}
\begin{document}

\title{Beyond Correlation: Evaluating Multimedia Quality Models with the Constrained Concordance Index}

\author{Alessandro Ragano, Helard Becerra Martinez, and Andrew Hines \IEEEmembership{Senior Member, IEEE}
\thanks{This publication has emanated from research conducted with the financial support of Science Foundation Ireland under Grant number 12/RC/2289\_P2. For the purpose of Open Access, the author has applied a CC BY public copyright licence to any Author Accepted Manuscript version arising from this submission. }
\thanks{Alessandro Ragano, Helard Becerra Martinez, and Andrew Hines are with the Insight Centre for Data Analytics and the School of Computer Science, University College Dublin, Dublin, Ireland (e-mail: alessandro.ragano@ucd.ie, helard.becerra@ucd.ie, andrew.hines@ucd.ie)}
}

\markboth{Journal of \LaTeX\ Class Files, Vol. 14, No. 8, August 2015}
{Shell \MakeLowercase{\textit{et al.}}: Bare Demo of IEEEtran.cls for IEEE Journals}
\maketitle

\begin{abstract}
This study investigates the evaluation of multimedia quality models, focusing on the inherent uncertainties in subjective Mean Opinion Score (MOS) ratings due to factors like rater inconsistency and bias. Traditional statistical measures such as Pearson's Correlation Coefficient (PCC), Spearman's Rank Correlation Coefficient (SRCC), and Kendall's Tau (KTAU) often fail to account for these uncertainties, leading to inaccuracies in model performance assessment. We introduce the Constrained Concordance Index (CCI), a novel metric designed to overcome the limitations of existing metrics by considering the statistical significance of MOS differences and excluding comparisons where MOS confidence intervals overlap. Through comprehensive experiments across various domains including speech and image quality assessment, we demonstrate that CCI provides a more robust and accurate evaluation of instrumental quality models, especially in scenarios of low sample sizes, rater group variability, and restriction of range. Our findings suggest that incorporating rater subjectivity and focusing on statistically significant pairs can significantly enhance the evaluation framework for multimedia quality prediction models. This work not only sheds light on the overlooked aspects of subjective rating uncertainties but also proposes a methodological advancement for more reliable and accurate quality model evaluation.
\end{abstract}


\IEEEpeerreviewmaketitle

\section{Introduction}
The evaluation of multimedia quality models~\cite{liu2018end,ragano2021more,freitas2018blind,korhonen2020blind,chen2020rirnet,chinen2020visqol,bianco2018use,akhtar2017audio,martinez2014no,ragano2023comparison}, is essential for assessing the efficacy of signal processing algorithms across various domains such as inpainting, enhancement, codecs, voice call standards, and media streaming. The performance of objective quality models is typically evaluated by comparing the predicted Mean Opinion Score (MOS) with human-rated MOS, employing statistical measures for accuracy. The MOS can be derived by averaging individual ratings on a 5-point Absolute Category Rating (ACR) scale, with the possibility of extending this scale to capture more nuanced user perceptions (e.g., a 10-point ACR scale).

Following the ITU-T P.1401 guidelines~\cite{ITUP1401}, the evaluation process involves statistical tools like the root mean squared error* (RMSE*), Pearson's correlation coefficient (PCC), and the outlier ratio (OR) to analyze the average error, linearity, and the prevalence of outliers, respectively. In addition, Spearman's rank correlation coefficient (SRCC) and Kendall's tau correlation coefficient (KTAU) offer insights into monotonic relationships and the balance between concordant and discordant pairs. These methodologies are widely used to evaluate regression tasks, but they neglect that subjective MOS ratings are highly influenced by several factors such as rating inconsistency, bias from MOS distribution~\cite{zielinski2008some}, order-effect, and rater-specific biases like culture and language. Ignoring this uncertainty can lead to inaccuracies in the performance evaluation of quality models. The exploration into evaluating instrumental quality models with a focus on rater uncertainty is limited, particularly neglecting factors such as the restriction of range, sample size, and variability among rater groups, which are critical for a comprehensive evaluation of multimedia quality prediction. These factors can be summarised as follows:

\begin{itemize}
\item Small \textit{sample size} impacts the robustness of statistical measures like PCC, SRCC, and KTAU, particularly when the evaluations are based on averaging across a limited set of conditions\footnote{We use the term condition to indicate a particular degradation at a specific intensity e.g., gaussian noise added to images at 20 dB signal-to-noise ratio (SNR)}. This can introduce a sampling bias, as these measures may not accurately reflect the population's characteristics with a small dataset, leading to potential inaccuracies in the evaluation of objective quality models~\cite{schonbrodt2013sample,de2016comparing}.
\item \textit{Restriction of range} refers to the issue arising when statistical metrics like PCC or SRCC are calculated over a subset of the full data range, leading to potential deviations in their values. This problem is accentuated in uncontrolled environments, such as user-generated content or in-the-wild assessments, where the quality range is uncontrolled and might not accurately represent the entire spectrum of quality~\cite{krasula2016accuracy,saad2014blind,li2023reqa}.
\item \textit{Variability among rater groups} emerges, especially in crowdsourced settings, because MOS is a relative scale as shown in several studies~\cite{van1995quality,mantiuk2012comparison,zielinski2008some,shirali2018mos,itu2016800}. In crowdsourcing ~\cite{tu2021ugc,wu2023exploring,li2021user,kancharla2021completely,zhang2021uncertainty,saha2023re,zhang2020learning,li2019quality}, different stimuli are assigned to different groups due to the large size of the dataset that needs to be annotated. As a consequence, every group will necessarily introduce a group bias effect. 
\end{itemize}

Krasula et al.'s work~\cite{krasula2016accuracy} marks a significant step forward in proposing a performance evaluation metric that identifies significantly different pairs and quantifies uncertainty in objective metric predictions, thus enhancing the evaluation framework to account for the statistical significance of subjective scores and the integration of data from varied subjective experiments. Another study that questions the reliability of PCC has been proposed for speech quality models~\cite{kolossa2016evaluating}. Here, the authors propose a Bayesian model selection to evaluate instrumental quality models, showing that PCC does not have high explanatory power compared to their approach. Both studies confirm the inadequateness of statistical metrics but do not address the three issues mentioned above: small sample sizes, restriction of range, and variability among rater groups. This paper seeks to address this gap by evaluating the performance of instrumental quality models under these three scenarios. We conduct a thorough examination of statistical metrics for instrumental quality models to evaluate their robustness in these scenarios.
In addition, we introduce the Constrained Concordance Index (CCI), a novel metric designed to assess the performance of instrumental quality models. CCI measures the capability of these models to accurately rank pairs where MOS has high precision and ignores the ones with uncertain MOS. Our findings demonstrate that typical statistical metrics (PCC, SRCC, and KTAU) lack robustness in the three scenarios mentioned above, whereas the proposed CCI effectively addresses this issue.

We proposed the CCI metric in our previous study to evaluate objective quality models for sound archives~\cite{ragano2023audio}\footnote{We introduced the CCI metric in \cite{ragano2023audio} with another name: pairwise ranking accuracy (PRA)}. In our previous paper~\cite{ragano2023audio} we have only used the CCI and motivated its usage but we did not compare it against traditional metrics and show its robustness for multimedia quality model evaluation.
In this paper, we slightly modify the CCI metric (see Section 3) and we do an extensive comparison of the CCI metric against traditional statistical metrics (PCC, SRCC, and KTAU) to evaluate the robustness against the three scenarios mentioned above. One of the contributions of this paper is also the evaluation of traditional statistical metrics in these three scenarios. 
To achieve this we select raters and samples from laboratory-based speech and image quality datasets through bootstrapping. By extracting a subset of stimuli and raters from these lab-based quality databases, we examine the extent to which statistical metrics diverge from their values obtained when using the full set of raters and stimuli.
In Section 2 we outline the motivations for proposing a new metric that overcomes the issues of traditional statistical metrics for the three studied scenarios: small sample size, restriction of range, and rater group variability. Section 3 is dedicated to the description of the proposed metric CCI. Experiment setup and results comparing all the statistical metrics are shown in Section 4 while a description of how to interpret and visualise the proposed CCI metric is described in Section 5. Finally, we provide a paper discussion in Section 6 and our conclusions in Section 7.
The Python code to reproduce the experiments and the CCI metric are available on GitHub\footnote{\url{https://github.com/alessandroragano/muqeval}}.
\section{Motivations}
\label{sec:motivations}
Current methods for evaluating instrumental quality models often overlook the uncertainty introduced by rater variability, which can lead to inaccurate assessments. It is a well-established fact that the 95\% confidence intervals (CIs) for MOS become wider as the number of raters decreases. For instance, with 24 raters, the 95\% CI may range between 0.5 and 0.7. However, with only 6 raters, this interval can expand to more than 1.5~\cite{pinson2022no}. This highlights the need to consider additional statistics beyond a single-point MOS estimate.

\begin{figure}[!b]
  \centering
  \includegraphics[width=0.90\linewidth]{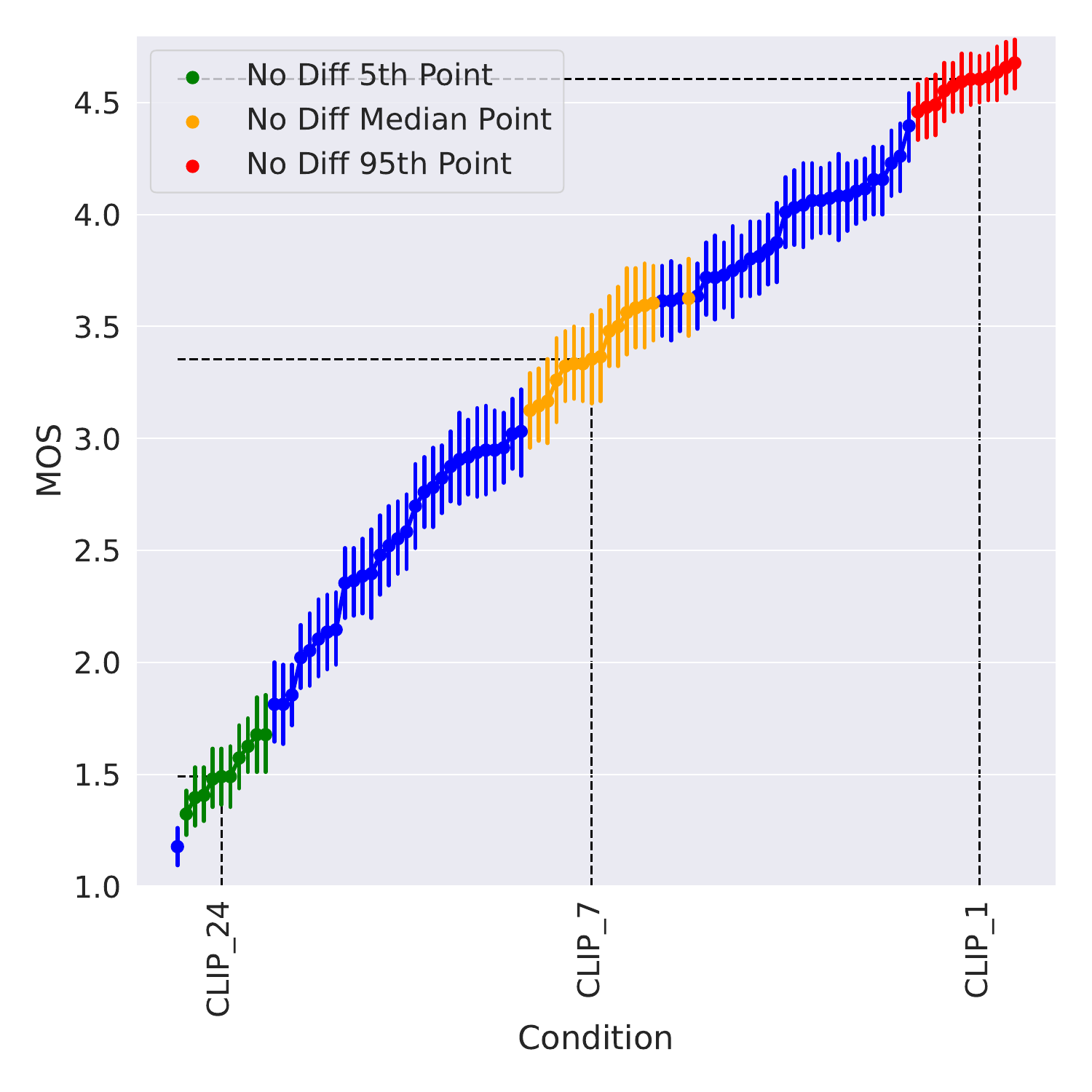}
  \caption{Statistically difference between 5th (Green), 95th (Red), and median (Orange) MOS condition with respect to all the other conditions.}
  \label{fig:tcdvoip_stats}
\end{figure}

Relying solely on MOS to measure QoE has been questioned due to its limited representation of QoE compared to other statistics such as standard deviation, confidence interval, and kurtosis~\cite{streijl2016mean}. Research on subjective methods has explored modelling rater subjectivity. Ho{\ss}feld et al. proposed using the standard deviation of opinion scores (SOS)~\cite{hossfeld2011sos}, and Seufert demonstrated the benefits of using MOS distributions instead of point estimate MOS values~\cite{seufert2019fundamental}. Similarly, instrumental quality models have shown improvements when predicting individual rater scores~\cite{chinen2021marginal}, modelling preference scores\cite{hu23d_interspeech}, or incorporating MOS distributions into the prediction model~\cite{faridee22_interspeech,gao2022image,liu2019comprehensive}. These studies support the notion that incorporating additional statistics or modelling rater subjectivity enhances QoE measurement for both subjective tests and instrumental quality models.

\begin{figure*}[!t]
	\centering
	\begin{subfigure}{0.3\textwidth}
            \centering
		\includegraphics[width=\linewidth]{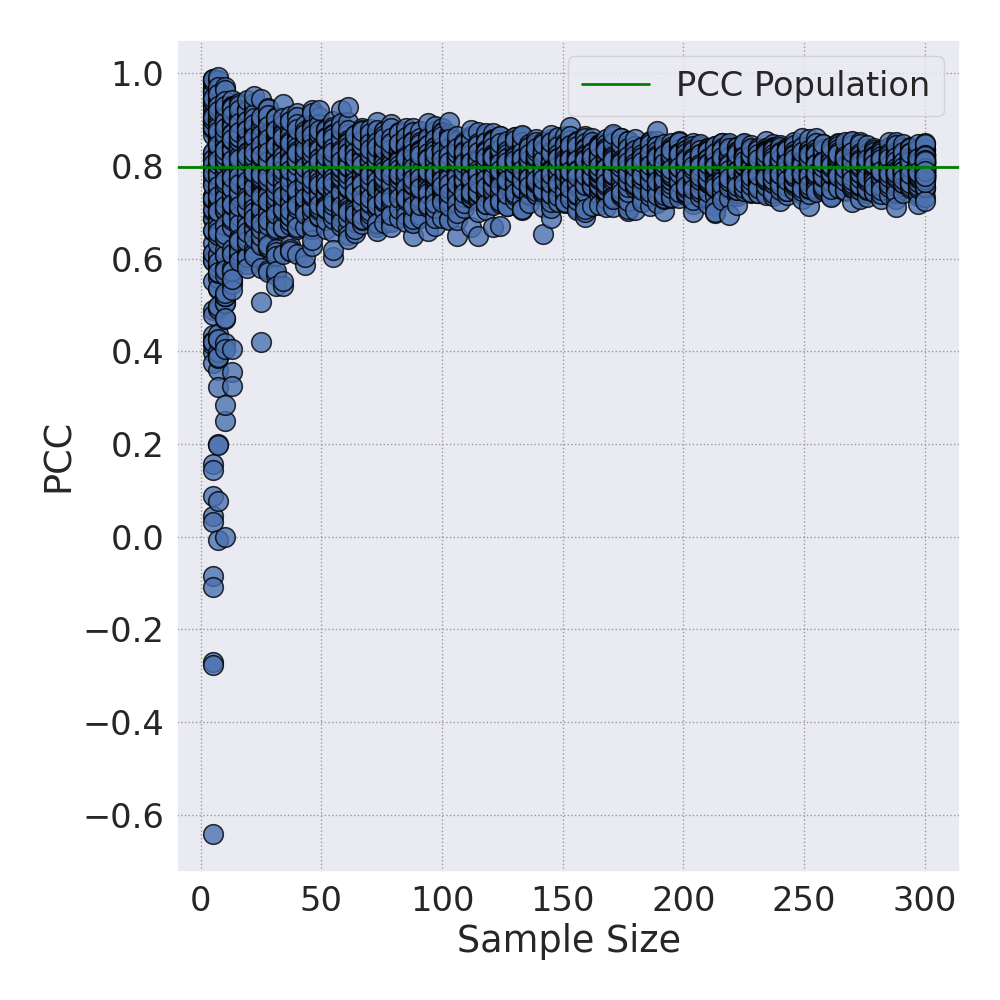}
		\caption{PCC}
		\label{fig:pcc_dev}
	\end{subfigure}
	\begin{subfigure}{0.3\textwidth}
		\includegraphics[width=\linewidth]{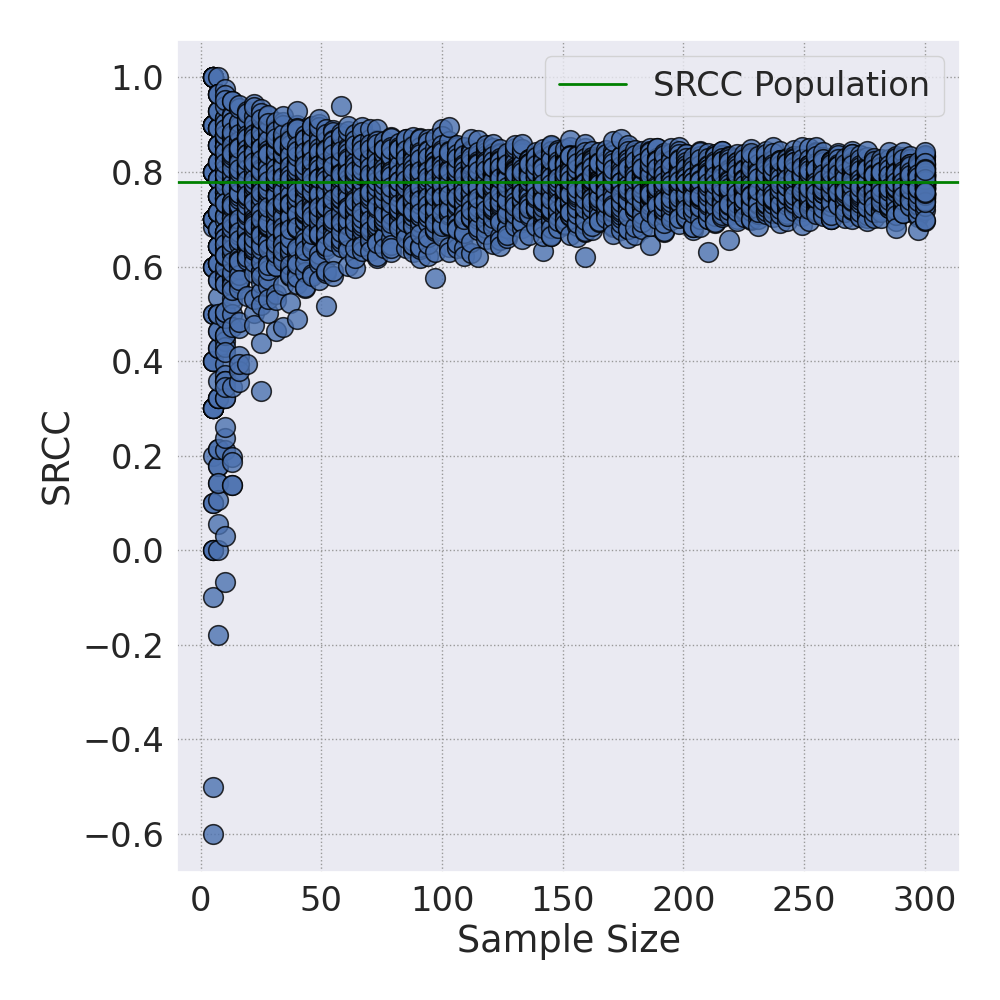}
		\caption{SRCC}
		\label{fig:srcc_dev}
	\end{subfigure}
	\begin{subfigure}{0.3\textwidth}
	        \includegraphics[width=\linewidth]{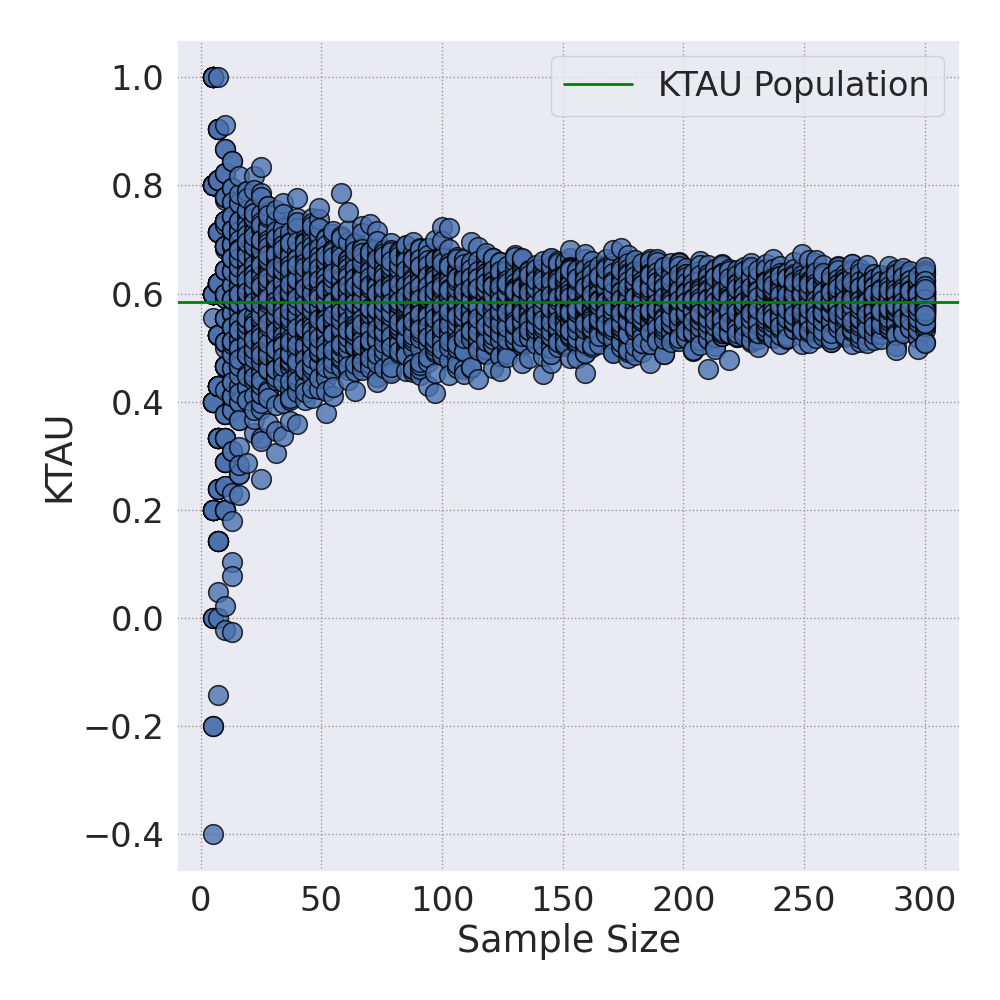}
	        \caption{KTAU}
	        \label{fig:ktau_dev}
         \end{subfigure}
	\caption{Deviation from the population of (a) PCC (b) SRCC (c) KTAU using simulated data with PCC=0.80, 1000 population data points,  and 100 random combinations of data points for each sample size.}
	\label{fig:small_sample_size}
\end{figure*}

Recent objective quality models based on deep learning are commonly trained and tested on MOS collected from ACR scores obtained with crowdsourcing. The ACR scale is easier to understand for non-expert raters (e.g., casual raters not working in the multimedia field) and allows the collection of large amounts of labels quicker than other methods such as pairwise comparisons.
One problem of ACR is that discrimination power can be lost for stimuli that are close in quality. 

This concept can be demonstrated using an example. Speech quality conditions subjectively rated with 24 raters for 4 samples per condition from the TCD-VoIP speech database~\cite{harte2015tcd} are illustrated in Figure \ref{fig:tcdvoip_stats}. The TCD-VoIP database is a speech quality database made of 96 conditions created from 5 degradations: clip, background noise, chopped speech, competing speaker, and echo.
The figure highlights the relationships between three speech quality conditions corresponding to the 5th (CLIP 24), 50th (CLIP 7), and 95th (CLIP 1) percentile rankings within the MOS distribution and their respective adjacent points. The adjacent points coloured in the figure, correspond to the non-statistically different conditions with respect to each chosen condition: red for CLIP 1, orange for CLIP 7, and green for CLIP 24. To identify conditions that did not show significant differences at these three points, we conducted the Wilcoxon signed-rank test, a non-parametric method for paired comparisons, complemented with a post-hoc correction for multiple tests.

Our observation indicates that various conditions exhibit no MOS quality differences across the three selected points and their adjacent points, and this pattern persists regardless of the condition examined. This phenomenon might stem from either an actual absence of quality difference among these conditions or the ACR scale's insufficient sensitivity to detect such differences. Therefore, it highlights the need for an evaluation metric designed to capture this level of uncertainty, rather than penalising objective models if they do not distinguish these subtle variations.

Below, we motivate why reduced sample sizes, variability in rater sampling, and range restriction are overlooked problems when evaluating instrumental quality model performance. We also explain how the inclusion of statistical measures like pair uncertainty can address and alleviate these issues by solving the problem evidenced by the scenario depicted in Figure~\ref{fig:tcdvoip_stats}.

\subsection{Sample Size}
\begin{table}
  \caption{Sample Sizes of Quality Databases}
  \label{tab:samplesize}
  \begin{tabular}{lccc}
    \toprule
    Database & Files & Conditions & Media \\
    \midrule
    TCD-VoIP~\cite{harte2015tcd} & 384 & 96 & Speech \\
    P.Suppl23 EXP1~\cite{sup232004itu} & 176 & 44  & Speech \\
    P.Suppl23 EXP3~\cite{sup232004itu} & 216 & 50  & Speech \\
    NISQA\_TEST\_FOR~\cite{mittag21_interspeech} & 240 & 60  & Speech \\
    NISQA\_TEST\_LIVETALK~\cite{mittag21_interspeech}  & 232 & 58 & Speech \\
    NISQA\_TEST\_P501~\cite{mittag21_interspeech} &240 & 60 & Speech \\
    JPEG XR~\cite{de2009subjective} & 186 & 30 & Image \\
    CSIQ~\cite{larson2010most} & 866 & 29 & Image \\
    IVC~\cite{ivc} & 185 & 25 & Image \\
    TID 2013~\cite{ponomarenko2015image} & 3000 & 124 & Image \\
  \bottomrule
\end{tabular}
\end{table}
In evaluating objective quality models, it is standard practice to calculate the average predictions and subjective MOS across stimuli under identical conditions, such as common encoding levels (e.g. mp3 at 96~kbps for audio), environmental SNR levels for speech (10~dB SNR babble) or network loss conditions. This averaging is aimed at minimizing the influence of confounded factors on human ratings. For instance, given the same condition, raters might prefer a speaker to another and consistently give slightly different rates. The same applies to images or videos, where content preferences affect ratings. Consequently, the number of conditions often serves as the effective sample size\footnote{In this paper we use the term sample size to indicate the size of the dataset used to calculate the coefficients. This can be either the number of conditions or the number of multimedia files in the dataset, based on how it is specified.} for calculating PCC or SRCC.

It is well-established that PCC, SRCC, and KTAU are vulnerable to inaccuracies when applied to datasets with low sample sizes~\cite{schonbrodt2013sample,de2016comparing}. This issue is demonstrated in Figure~\ref{fig:small_sample_size} where we illustrate the discrepancy between coefficients derived from smaller sample sizes versus those from the population, using simulated data with a PCC of 0.80. 
Simulated data are created by sampling 1000 data points from a multivariate normal distribution. In Figure~\ref{fig:small_sample_size} we extract 100 combinations of data points at each sample size.
Stability in these coefficients typically emerges with sample sizes of around 100 or more. As evidenced in Table~\ref{tab:samplesize}, many databases commonly used in multimedia quality evaluation do not reach this threshold. We observe that several quality databases used for testing performances often consist of as few as 30 or 40 different conditions. Some databases such as KADID10k~\cite{lin2019kadid} include more conditions, but they are built using noisy crowdsourcing labels and serve mostly for training purposes given their size.  We notice that evaluating per condition is the recommended approach in multimedia quality models to cancel out individual differences due to content preferences. This introduces a significant risk of sampling bias if the goal is to make significant conclusions when comparing quality models, leading to potential deviations from true population metric values. Despite their widespread use, SRCC and PCC's low accuracy with smaller sample sizes is often overlooked in the evaluation of instrumental models. This variability highlights the necessity for a metric robust to sample size variations, which is essential for evaluating the generalizability of quality models across diverse conditions. 

\subsection{Rater Variability}
As the field of no-reference media quality assessment evolves with the integration of supervised deep learning techniques, the practice of crowdsourcing raters for both training and testing datasets has become increasingly prevalent. This method involves participants rating the quality of media via web applications, diverging from controlled experimental settings. In such uncontrolled environments, variability arises from participants using different equipment and being subjected to varying conditions, which may lead to inconsistencies in ratings due to potential distractions or incentives to cheat. This inherent variability necessitates a larger pool of participants to achieve statistical power comparable to traditional lab-based studies. The reliance on deep learning for objective quality models further compounds this issue, requiring extensive data labelling across multiple rater groups to train and validate models effectively. Indeed, it is unfeasible to label a large dataset by assigning the same stimuli to all the raters. 

The problem with rater groups in the same dataset is that MOS is a relative scale, as well-documented in many studies, e.g. \cite{van1995quality,mantiuk2012comparison,zielinski2008some,shirali2018mos,itu2016800}. As a consequence, each rater group introduces a bias based on the specific stimuli they assess. Normalizing the quality scale across several rater groups presents a significant challenge due to this inherent variability~\cite{ragano20_interspeech}. The issue persists across both training and test datasets in crowdsourced environments, highlighting a gap in multimedia research. Despite the growing adoption of 'in-the-wild' crowdsourcing to minimise the domain gap between simulated and real-world data, the effect of group bias in deep learning-based quality models has been unexplored. This bias might occur in various datasets, including those focusing on user-generated content (UGC) video quality \cite{tu2021ugc,wu2023exploring,li2021user,kancharla2021completely}, in-the-wild image or video quality assessment \cite{zhang2021uncertainty,saha2023re,zhang2020learning,li2019quality}, and audio archive quality assessment \cite{ragano2023audio,ragano2022automatic}, all of which rely on crowdsourcing with multiple rater groups. 
In addressing the variability introduced by different rater groups, the use of anchor conditions in both training and rating sessions is used to align the perceptual and the rating scales~\cite{ITUP808}. 
However, this method does not fully account for the subjective biases influenced by individual content preferences and human bias. For example, the evolution of speech synthesis has caused a shift in MOS scores of previously rated text-to-speech systems~\cite{le2022back}, emphasizing how MOS is a relative scale even if the same anchors are used. It is not known whether the rater group effect might have a negative impact on statistical metrics. If high variability is observed, the development of a new metric resistant to such individual differences becomes necessary.   

In this paper, we conduct simulations to assess the robustness of traditional statistical metrics against rater variability. By comparing these findings with our newly proposed evaluation metric, this study is the first comprehensive examination of how statistical metrics and quality models respond to rater group bias.

\subsection{Restricted Range}
\begin{figure}[!t]
  \centering
  \includegraphics[width=0.90\linewidth]{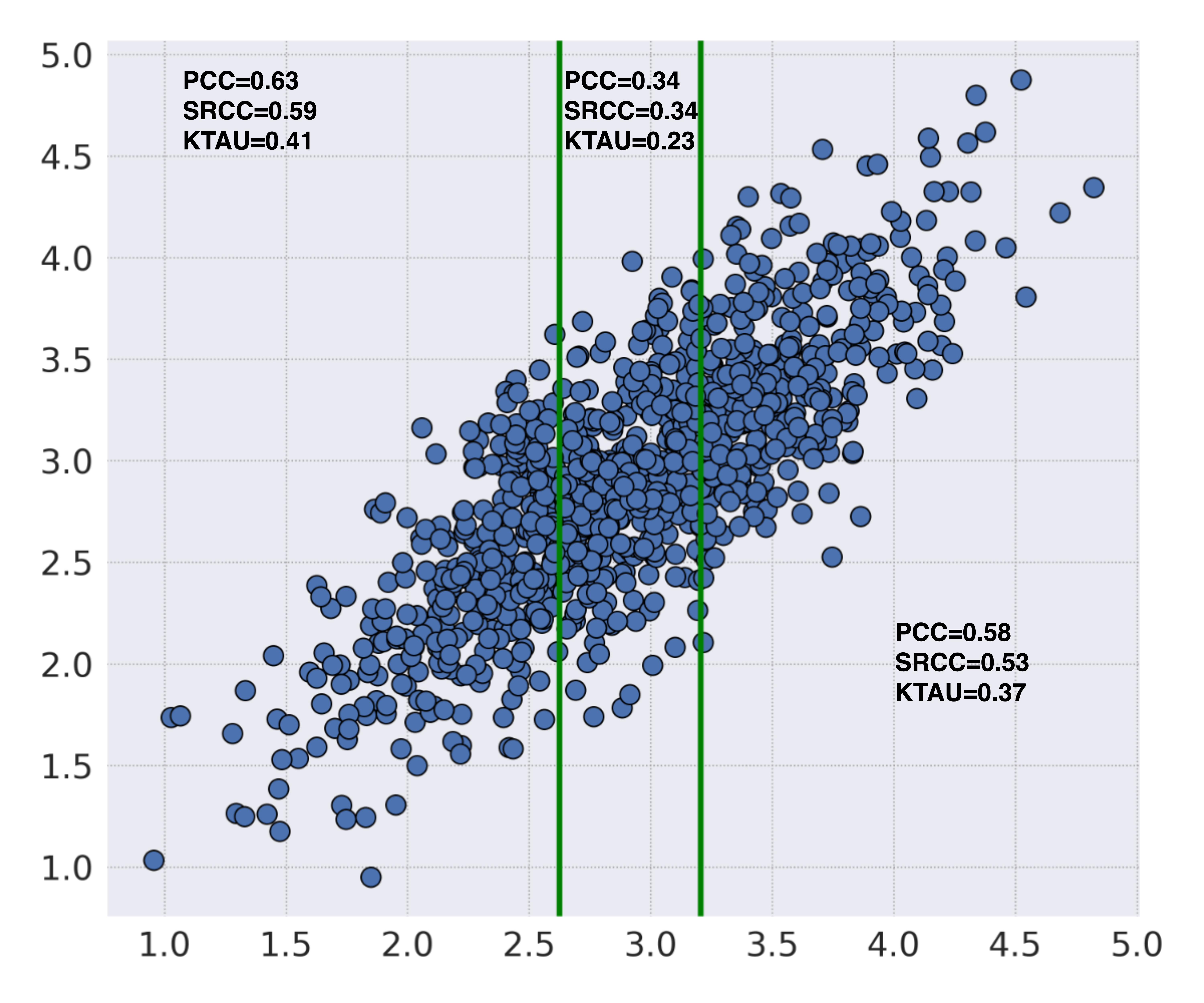}
  \caption{Simulated data with PCC=0.80. Subintervals are created by splitting the values on the x-axis variable into 3 regions with an equal number of points: [1, 2.65], [2.65, 3.20], [3.20, 4.82].}
  \label{fig:rr_simulation}
\end{figure}

The accuracy of PCC and SRCC is significantly compromised when applied to subsets of data, a phenomenon well-documented but insufficiently addressed in existing research~\cite{krasula2016accuracy,saad2014blind,li2023reqa}. This issue is shown in our simulation of 1000 data points with a baseline PCC of 0.80, revealing substantial deviations of these coefficients across various subintervals (Figure~\ref{fig:rr_simulation}). Such deviations highlight the susceptibility of these metrics to the 'restriction of range' effect, particularly problematic in evaluations involving 'in-the-wild' data collected under uncontrolled conditions, such as user-generated content (UGC) and in-the-wild image or video quality assessments~\cite{tu2021ugc,wu2023exploring,li2021user,kancharla2021completely,zhang2021uncertainty,saha2023re,zhang2020learning,li2019quality}.

In controlled environments, where degradation levels can be synthetically generated, the MOS range adequately reflects the scope of the domain within which one wants to develop a quality model. However, the uncertain quality range of in-the-wild data often leads to misleading outcomes when applying PCC or SRCC, as these coefficients might not accurately represent the objective model quality prediction performance over an arbitrarily restricted quality spectrum. This is especially true when quality assessments focus on a narrow selection of points, potentially yielding results that diverge significantly from those based on the full dataset~\cite{saad2014blind}.

The challenge of collecting a comprehensive quality range under uncontrolled conditions, such as with video UGC, highlights the limitations of current metrics in accurately reflecting performance~\cite{li2023reqa,ragano20_interspeech}. The specific quality range of a dataset, often unknown until after data collection, necessitates an evaluation metric less dependent on the span of quality levels present.

\subsection{A New Performance Evaluation Metric}
Motivated by the evidence regarding the influence of reduced sample sizes, variability in rater sampling, and range restriction when comparing instrumental quality models, we propose the new metric CCI to address all the problems above. CCI is based on the principle of excluding comparisons where MOS difference is not statistically significant, thereby ensuring that only meaningful differences in quality are considered. 

This approach aims to provide a statistical metric for more reliable performance evaluation for objective quality models, particularly useful in studies with limited small sample sizes. By focusing on statistically significant pairs, CCI minimizes sampling bias and ensures evaluations reflect conditions representative of a broader, yet inaccessible, population. Traditional metrics give the same importance to every stimulus in the dataset. This will skew the evaluation in small sample-size scenarios where uncertain stimuli will have more influence. CCI, by design, filters out these uncertain pairs, thus preventing their impact on the overall assessment. This methodology ensures that instrumental quality models are evaluated on pairs where MOS has high statistical power.

Additionally, CCI tackles the issue of group bias by disregarding uncertain pairs, which are often influenced by individual content or speaker preferences. We assume that such bias predominantly affects pairs with lower inter-rater agreement (and therefore higher confidence intervals), where quality perceptions are more subjective. Uncertain pairs are the ones whose ranking has a higher chance of getting swapped if another group of raters is used. Ignoring these uncertain pairs, therefore, will help to reduce variability among different rater groups, leading to a more accurate assessment of model performance.

Moreover, CCI addresses the restricted range effect by considering the uncertainty inherent in user ratings. Traditional metrics can be misleading when evaluating quality within narrow ranges, such as those categorized within 'Bad' or 'Excellent' regions (Figure \ref{fig:PESQ_rr}). CCI mitigates this by excluding data points that lack statistical significance, thereby refining the evaluation process to include only meaningful comparisons.

\begin{figure*}[!b]
	\centering
	\begin{subfigure}{0.32\textwidth}
            \centering
		\includegraphics[width=\linewidth]{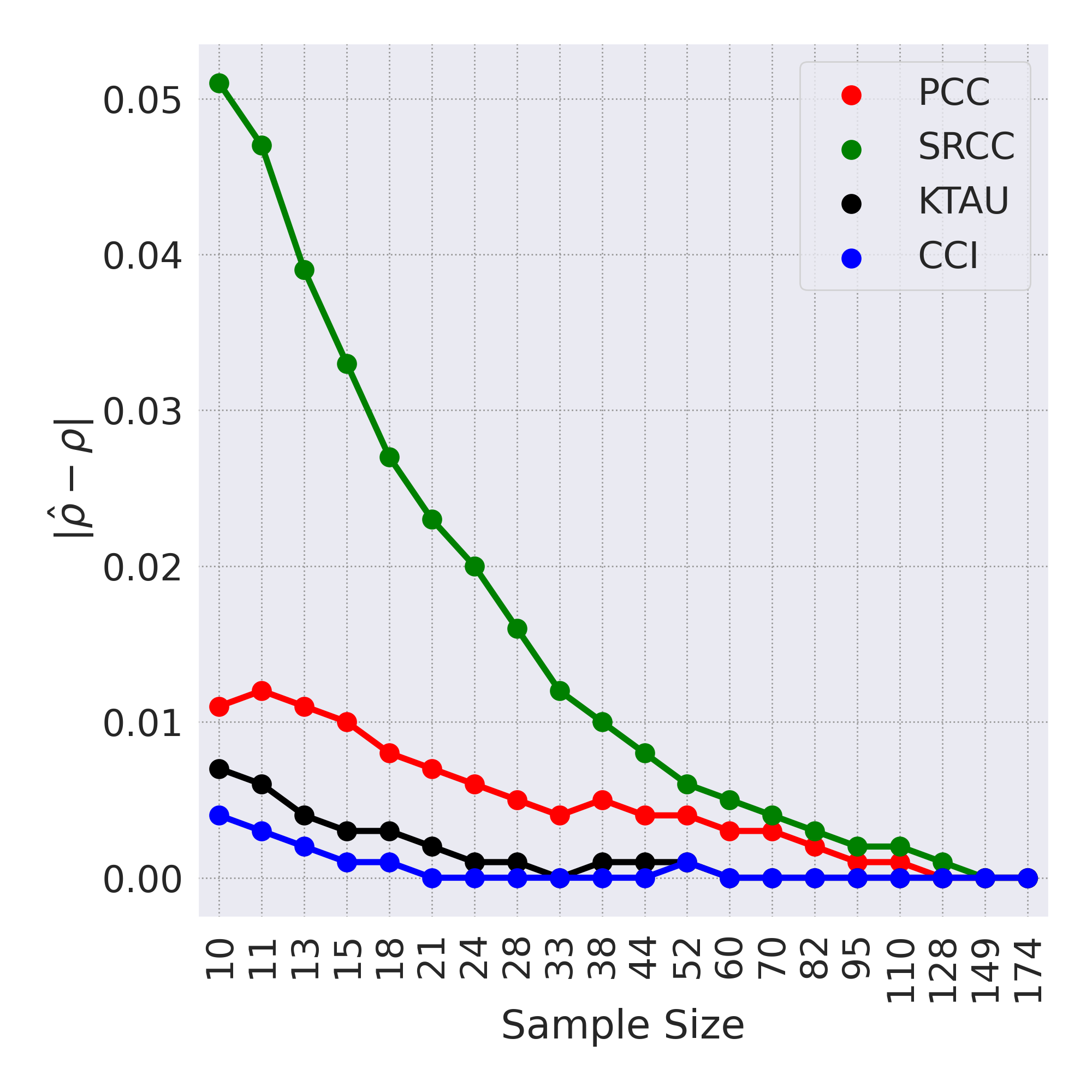}
		\caption{Sample Mean}
		\label{fig:exp1_pesq_sample_mean}
	\end{subfigure}
	\begin{subfigure}{0.32\textwidth}
		\includegraphics[width=\linewidth]{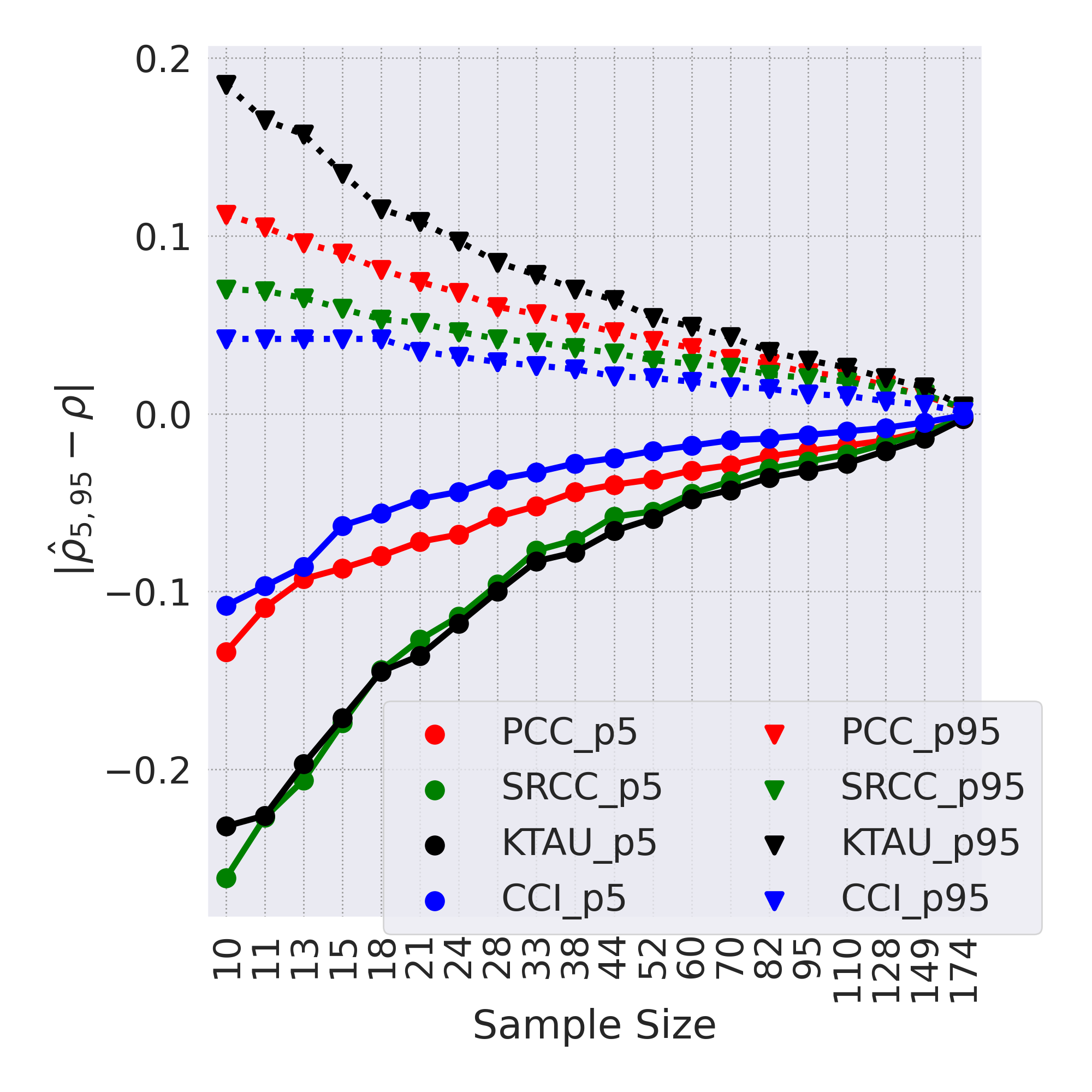}
		\caption{5th and 95th Percentile Deviation}
		\label{fig:exp1_pesq_percepoint}
	\end{subfigure}
	\begin{subfigure}{0.32\textwidth}
	        \includegraphics[width=\linewidth]{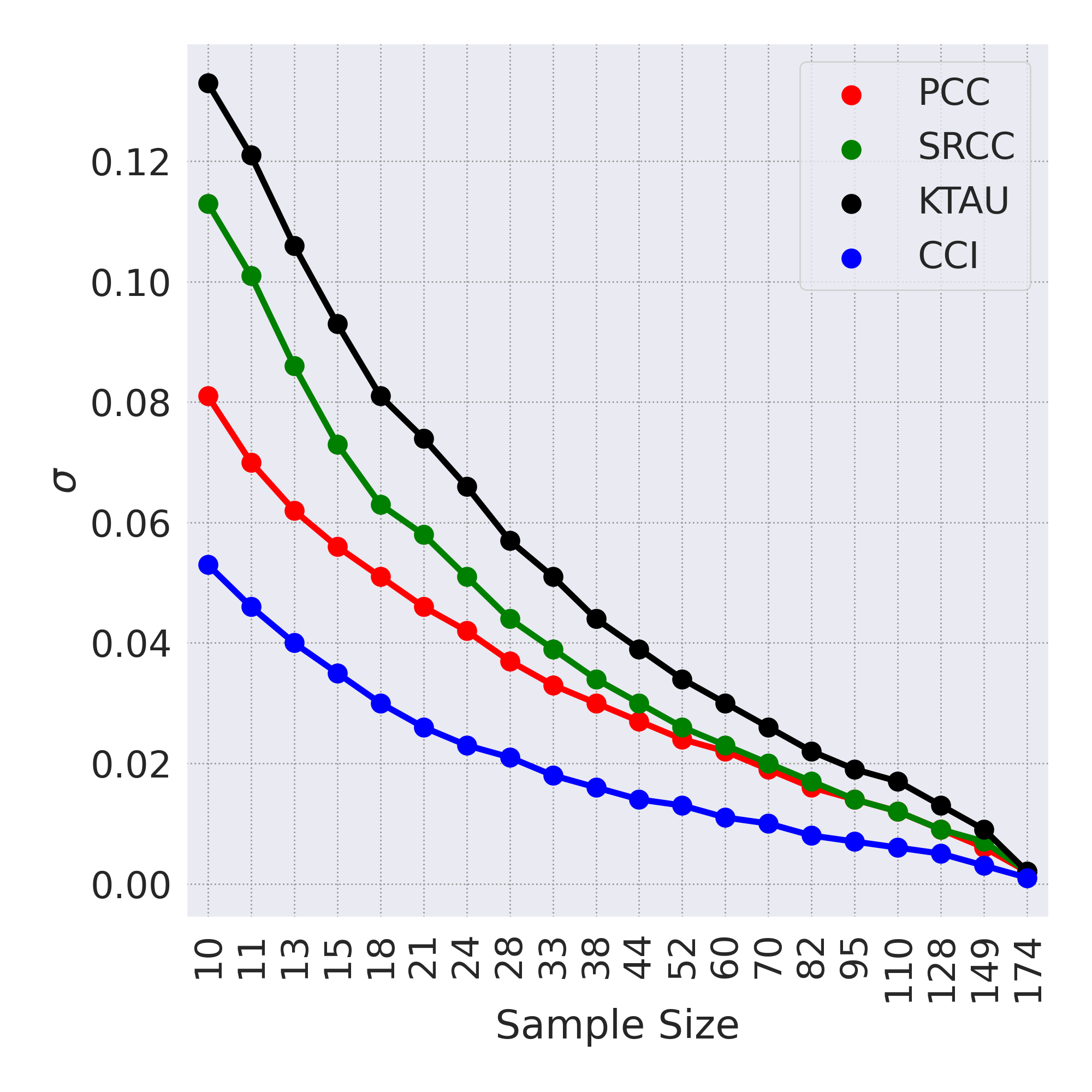}
	        \caption{Standard Deviation}
	        \label{fig:exp1_pesq_std}
         \end{subfigure}
	\caption{(a) Sample mean, (b) 5th and 95th percentiles, and (c) standard deviation, using PESQ  calculated on P23 EXP1.} 
	\label{fig:exp1_pesq}
\end{figure*}

\section{Constrained Concordance Index}
\label{sec:CCI}
We describe our proposed evaluation metric CCI in this section and then compare it with other statistical metrics in the next section.

Let us consider a dataset
$$\mathcal{D}=\{(\bm{x}_1,y_1), (\bm{x}_2, y_2), ... 
(\bm{x}_N,y_N)\}$$
where $\bm{x}_i$ are speech or images and $y_i \in R$ is the target variable MOS which can be obtained as $$y_i=\frac{1}{M}\sum\limits_{m=1}^{M} R_m$$ where $M$ is the number of raters and $R_m$ is the individual rating of the $m$-th rater for the stimulus $i$. Let us define the predicted MOS values which are the output of objective quality models as $\hat{y}_i$.

Let us consider $$\mathcal{S} = \{(a,b) | a, b \in \mathcal{D}, |y_a - y_b|>\tau\}$$ the set of all the combinations of $\mathcal{D}$ subject to the constraint $|y_a - y_b| > \tau$. The evaluation metric is defined as: 

\begin{equation}
CCI=\frac{1}{|S|}\sum_{a<b}\frac{\sign(y_a-y_b)\sign(\hat{y}_a-\hat{y}_b) + 1}{2}, \forall a,b \in \mathcal{S}.
\end{equation}

The $\sign$\footnote{Notice that the $\sign$ function is usually equal to 0 if the dependent variable is 0 e.g., $y_a=y_b$. In our definition, we exclude this scenario since the pairs with equal MOS are not considered valid pairs because they are not statistically different.} function is defined as follows:
\begin{equation}
  \sign(y_a - y_b)=
  \begin{cases}
    1, & \text{if}\ y_a>y_b \\
    -1, & \text{if}\ y_a<y_b.
  \end{cases}
\end{equation}
The CCI represents the number of concordant pairs (i.e. where both predictions and subjective MOS agree on ranking pairs) over the total number of pairs with the constraint that not all the pairs should be used. Concordant pairs are computed as follows. Given two stimuli $a$ and $b$, the rank of the predictions agrees with the rank of the MOS when either both \mbox{$\sign(y_a-y_b)$} and \mbox{$\sign(\hat{y}_a-\hat{y}_b)$} are equal to $1$ or equal to $-1$. Summing the constant $1$ in the numerator makes sure that the discordant pairs are not counted e.g., $\sign(y_a-y_b)=-1$ and $\sign(\hat{y}_a-\hat{y}_b)=1$. It must be noted that adding the constant will shift the CCI in the range 1-2. To get a range from 0 to 1, we divide the numerator by 2. 

The CCI considers only the pairs $(a,b)$ whose MOS 95\% confidence intervals ($CI(\cdot)_{95}$) do not overlap. The threshold for constraining the concordance index is set to

\begin{equation}
\label{eq:tau}
  \tau=\frac{CI(a)_{95}}{2}+\frac{CI(b)_{95}}{2}.
\end{equation}

The significance of the confidence interval is chosen as $95\%$ which is a typical recommended value for the MOS ratings~\cite{ITUP1401}. 

To understand whether two stimuli or conditions are statistically different we can inspect confidence intervals and conclude that: 1) if they overlap, we do not know whether there is or not a statistical difference and a statistical test is required.; 2) if they do not overlap, there is no need to run a statistical test since the two stimuli or conditions are statistically different. This means that by taking non-overlapping pairs, we use a conservative approach with the advantage of avoiding a statistical test which presents several limitations. The first issue is that statistical tests for all the pairs are time-consuming. The second issue lies in the fact that using tests to evaluate statistical significance between pairs can violate normality assumptions of the individual ratings as long as there is a sufficient number of raters as discussed in~\cite{narwaria2018data}. This still presents a limitation as we are addressing in-the-wild scenarios where the number of raters could be low as in the case of large datasets labelled with crowdsourcing. The third problem is that every statistical test gives different results and test assumptions should be carefully checked on every dataset.

In our previous work~\cite{ragano2023audio}, the CCI metric was different in the calculation of the constraint. Instead of considering as valid pairs, the ones where confidence intervals do not overlap (Equation \ref{eq:tau}), we used the average confidence interval of the training set as a valid MOS distance between two stimuli i.e., $|y_a - y_b| > \tau$ where $\tau=\frac{1}{N}\sum_{i=1}^{N}CI(i)_{95}$. However, this approach can be limiting as not all the confidence intervals of the training/development set might not be available. Also, some quality models such as PESQ are not data-driven. 

Confidence intervals can be either obtained from the individual ratings or the conditions, depending on which scenario is evaluated. The main difference between the two approaches is in the calculation of the unbiased standard deviation.
In CCI, confidence interval of the $i$-th stimulus is calculated as follows:

\begin{equation}
CI(i)_{95} = t(0.05, M)\frac{\lambda}{M}
\end{equation}

where $t(0.05, M)$ is the $t$-value obtained from the $t$-distribution, $\lambda$ is the unbiased standard deviation, and $M$ is the number of raters. 
The $t$-value should be found on the lookup table if the number of raters $M$ is lower than $30$ otherwise $1.96$ could be used. The unbiased standard deviation is calculated differently for file-based CCI or condition-based CCI as recommended in the ITU-T Rec. P.1401~\cite{ITUP1401}. 
For a file-based assessment, the standard deviation is given by:
\begin{equation}
    \sigma_{j,i} = \sqrt{\frac{\sum\limits_{m=1}^{M} (x_{j,i,m} - y_{j,i})^2}{M-1}} 
\end{equation}
where $m=1 \dots M$ are the raters, $i=1 \dots I$ are the files in the dataset, $j=1 \dots J$ are the conditions, $x_{j,i,m}$ is the individual ACR score assigned by the $m$-th rater to the $i$-th file corrupted with condition $j$.
If performances are evaluated per condition, the standard deviation for each condition is calculated as:
\begin{equation}
    \sigma_{j} = \sqrt{\frac{M-1}{N-1}\sum\limits_{i=1}^{I}\sigma_{j,i}^2}
\end{equation}
where N is the total number of votes for each condition $j$.

Like other statistical metrics, the CCI range goes from 0 to 1.  The highest CCI value is obtained when the model correctly ranks all the pairs $(i,j)$ in the constrained set $\mathcal{S}$. In the next section, we run simulations that show the CCI metric is more robust than PCC, SRCC, and KTAU in the three scenarios studied in this paper.

\begin{figure*}[!t]
	\centering
	\begin{subfigure}{0.32\textwidth}
            \centering
		\includegraphics[width=\linewidth]{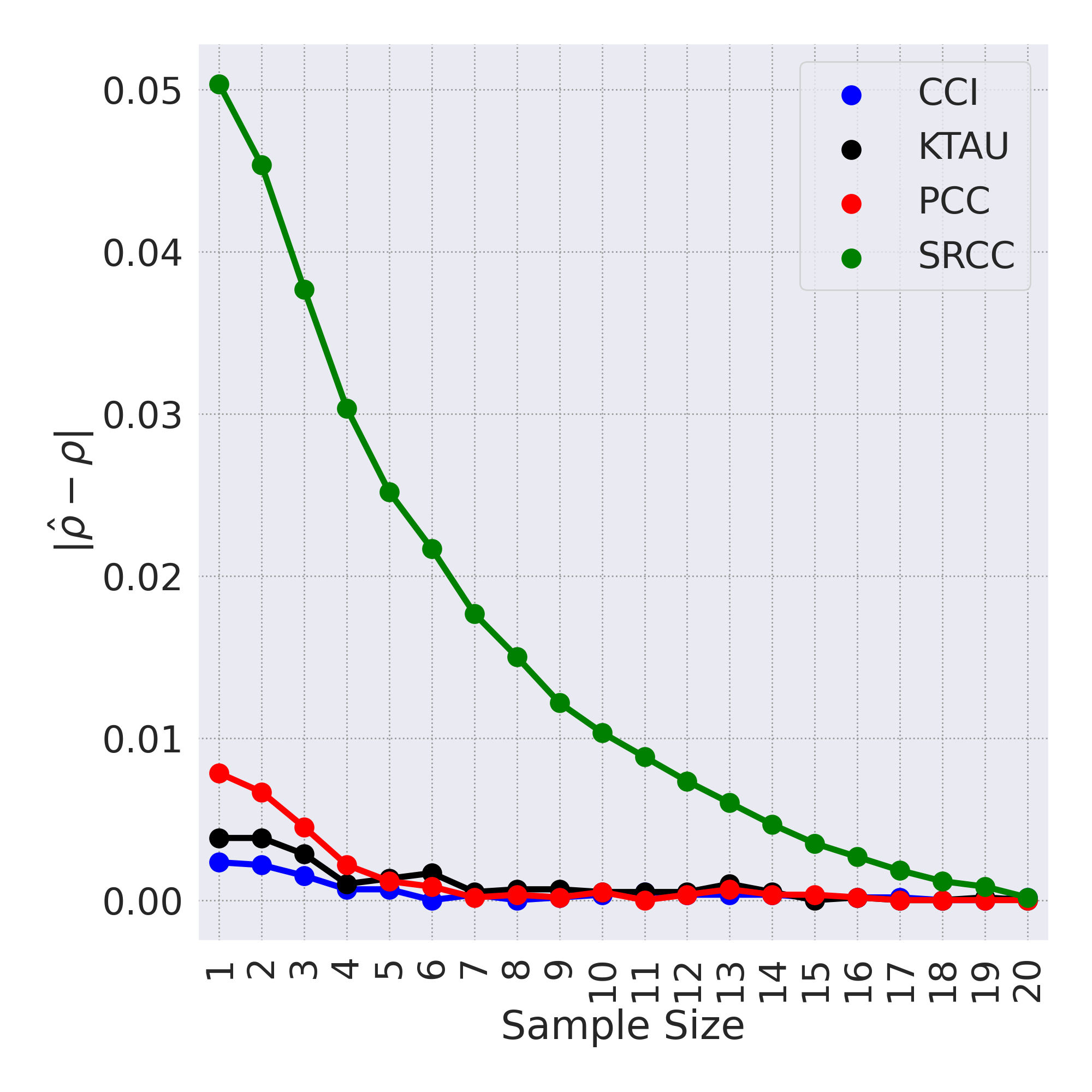}
		\caption{Sample Mean}
		\label{fig:exp1_sample_mean}
	\end{subfigure}
	\begin{subfigure}{0.32\textwidth}
		\includegraphics[width=\linewidth]{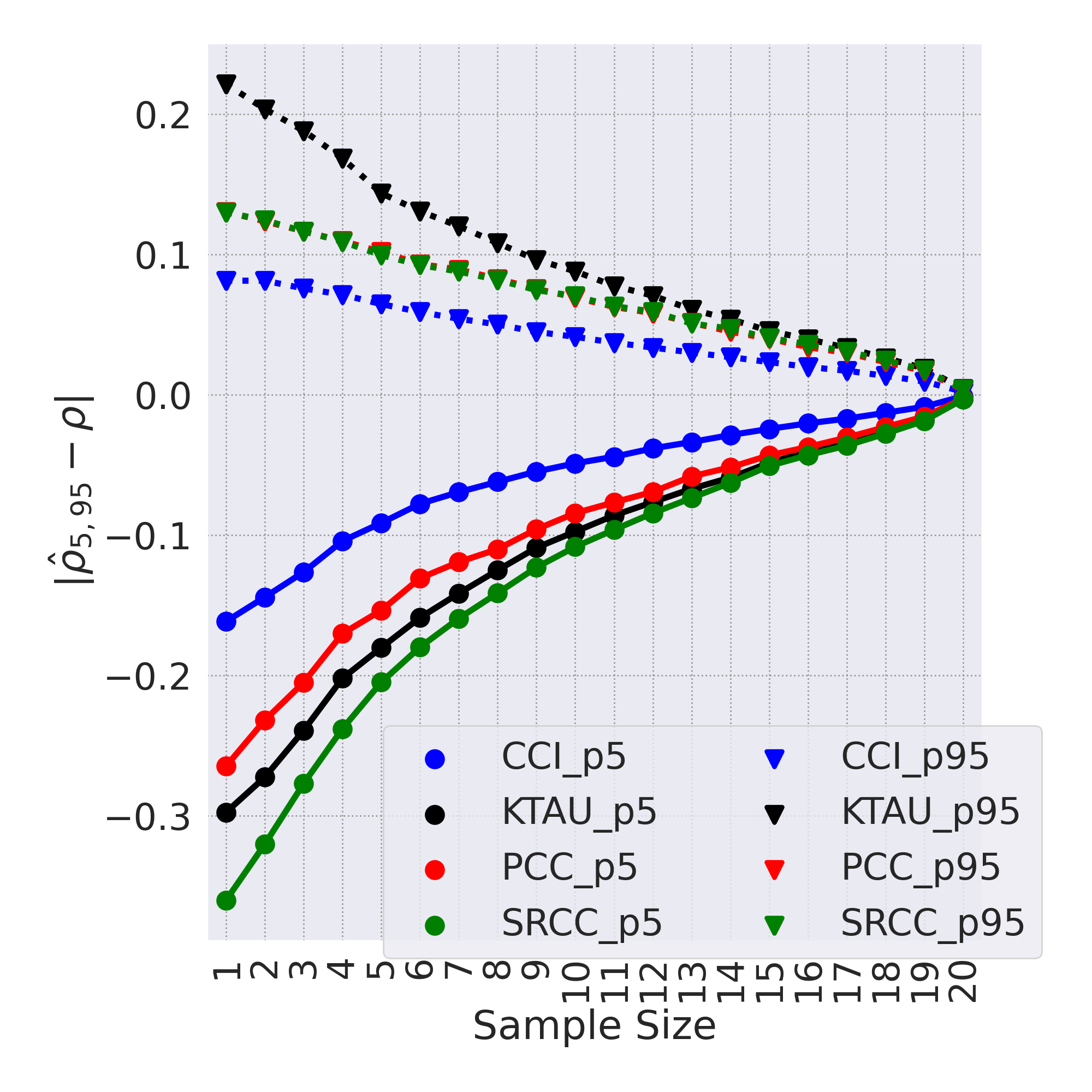}
		\caption{5th and 95th Percentile Deviation}
		\label{fig:exp1_percepoint}
	\end{subfigure}
	\begin{subfigure}{0.32\textwidth}
	        \includegraphics[width=\linewidth]{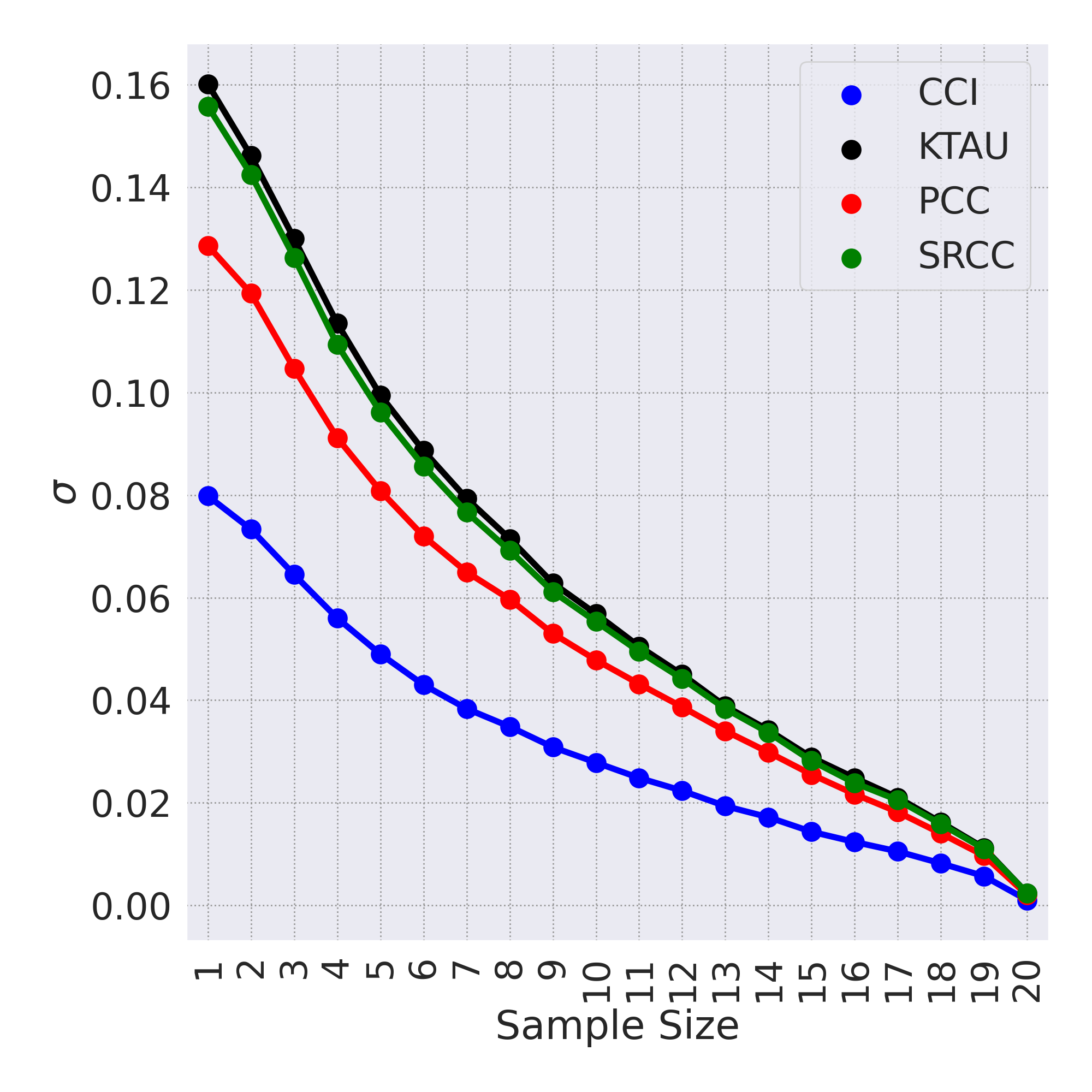}
	        \caption{Standard Deviation}
	        \label{fig:exp1_std}
         \end{subfigure}
	\caption{Aggregated (mean) of (a) sample mean, (b) 5th and 95th percentiles, and (c) standard deviation, using PESQ and VISQOL computed on TCD VOIP, P23 EXP1 and P23 EXP3. Objective quality scores are averaged on the sample index. The latter corresponds to different sample sizes according to which database is used. To retrieve the real sample size of each database using the sample index, see the mapping in Table \ref{tab:samplesizemapping}.} 
	\label{fig:exp1_aggregated}
\end{figure*}

\begin{table*}
  \caption{Sample Size Index Mapping Used in Figure \ref{fig:exp1_aggregated}}
  \label{tab:samplesizemapping}
  \ra{0.5}
  \centering
  \begin{tabular}{llllllllllllllllllllllllllllllllllllllllllllllllllllllllllllllllllllllllllllllllll}
    \toprule
    Index & 1 & 2 & 3 & 4 & 5 & 6 & 7 & 8 & 9 & 10 & 11 & 12 & 13 & 14 & 15 & 16 & 17 & 18 & 19 & 20 \\
    \midrule
    P23 EXP1 & 10 & 11 & 13 & 15 & 18 & 21 & 24 & 28 & 33 & 38 & 44 & 52 & 60 & 70 & 82 & 95 & 110 & 128 & 149 & 174 \\
    P23 EXP3 & 10 & 11 & 13 & 16 & 19 & 22 & 26 & 30 & 36 & 42 & 50 & 58 & 69 & 81 & 95 & 112 & 131 & 155 & 182 & 214 \\
    TCD-VOIP & 10 & 12 & 14 & 17 & 21 & 26 & 31 & 38 & 46 & 56 & 68 & 82 & 99 & 120 & 146 & 177 & 214 & 260 & 315 & 382 \\
    
  \bottomrule
\end{tabular}
\end{table*}

\section{Experiments}
\label{sec:experiments}
In our experiments, we first conduct an evaluation on speech media to demonstrate the characteristics of CCI in detail. Next, we show that these characteristics generalise to other quality models for other domains with an image quality example.
For the speech domain, we use 2 speech quality models, PESQ~\cite{rix2001perceptual} and ViSQOL~\cite{chinen2020visqol}.
PESQ is a full-reference speech metric that has been standardized (ITU P.862) and works for both narrowband speech (8 kHz) and wideband speech (16 kHz). PESQ has been extensively used in the audio community but shows several pitfalls for new generative speech codecs~\cite{jassim2020speech,ragano2023nomad}. ViSQOL is a full-reference metric that works for both wideband and fullband (48 kHz) audio and has been shown to correlate better than PESQ for time-warped speech. 
The databases that we use are P23 EXP1, P23 EXP3, and TCD~VoIP. The P23 EXP1 is a database for various codecs while P23 EXP3 includes conditions made of one codec (8kbps) under various channel degradations. The TCD~VoIP database has been described above and includes several degradations that can be found in VoIP calls: clipping, background noise, chopped speech, echo, and competing speakers. 
For the image domain, we use structured similarity image metric (SSIM), which is a well-established full-reference image quality model~\cite{wang2004image}. We evaluated SSIM on the JPEG XR compression dataset, which contains compressed images generated using different image coding algorithms at various bitrate values. We only use one metric to evaluate whether the same approach can be applied to different media other than speech.

It must be noted that we conduct a file-based assessment because it allows us to calculate robust population values since more samples are used when calculating correlations. However, our experiments can easily be adapted to a condition-based assessment.

\subsection{Experiment 1: Sample Size}
In this experiment, we analyse the robustness of statistical metrics when evaluating the performance of quality models by varying the sample size. Given a MOS database $\mathcal{D}$ with $N$ stimuli, we evaluate the statistical metrics by taking 20 sample sizes logarithmically spaced until the maximum sample size minus 2 (e.g., 382 for TCD-VOIP). We remove two samples to calculate statistical metrics that closely approximate the population values. Intervals between sample sizes increase logarithmically since the differences in correlation coefficients are more significant at lower sample sizes. For example, the difference in PCC between sample sizes of 10 and 12 is greater than the difference between sample sizes of 160 and 162.
To evaluate the robustness of correlation measures, 1000 different subsets are randomly generated for each sample size. For example, we sample 1000 different combinations of 11 stimuli, then 1000 different combinations of 13 stimuli, and so on. We consider as a parameter population, the statistical metrics calculated using the original sample size in the database. Statistical metrics are calculated for each of the drawn subsets and three statistics are reported: the absolute difference between the sample mean and the population mean, the standard deviation that measures the robustness against random sampling, the difference between both 5th percentile and 95th percentile values and the population mean. 

\begin{figure}[!t]
  \centering
  \includegraphics[width=0.80\linewidth]{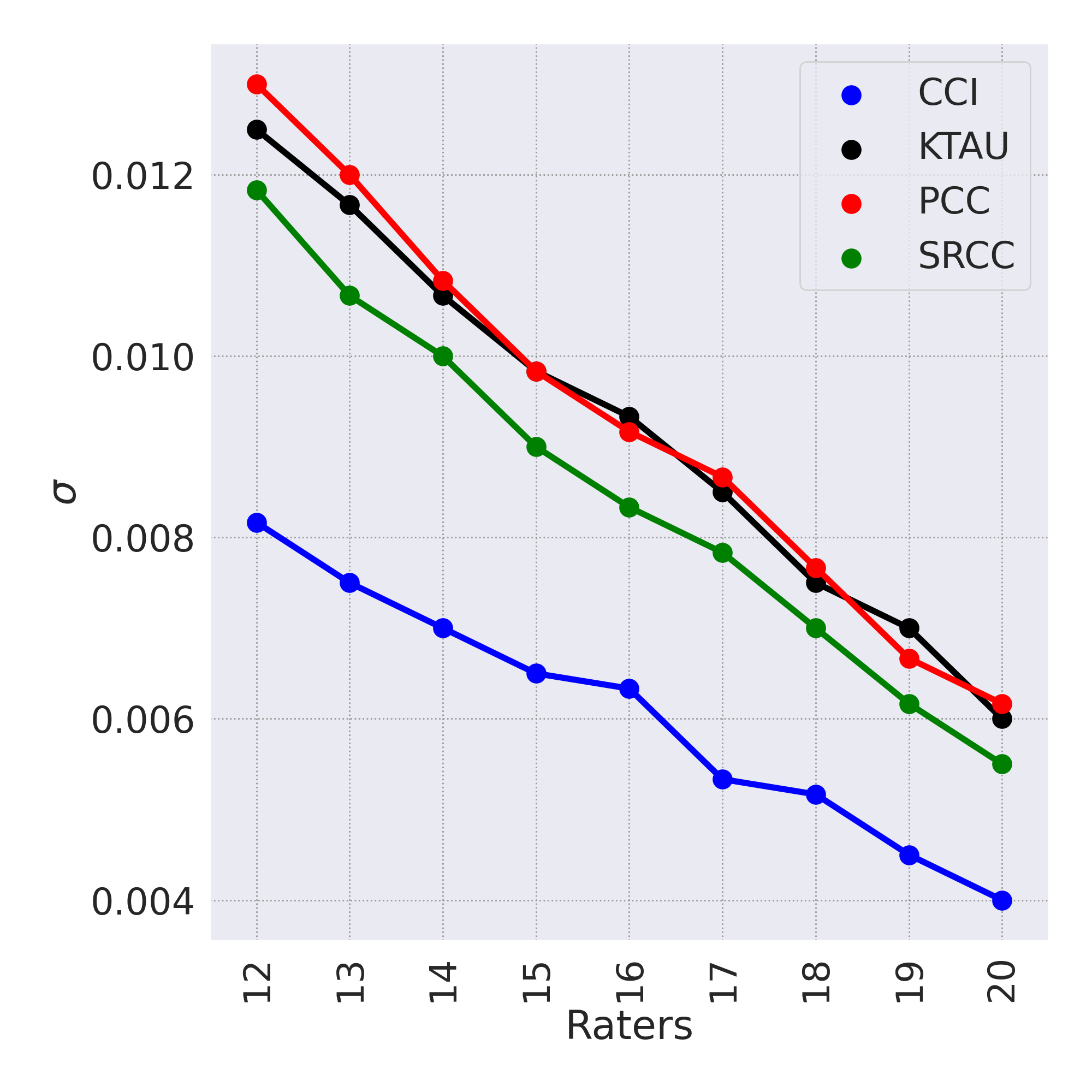}
  \caption{Standard deviation computed on 1000 subrater groups for each rater size. Statistical metrics calculated using VISQOL/PESQ predictions aggregated and MOS scores of the speech datasets P23 EXP1/P23 EXP3/TCD-VoIP aggregated}
  \label{fig:exp2_std}
\end{figure}

\begin{figure}[!t]
  \centering
  \includegraphics[width=0.80\linewidth]{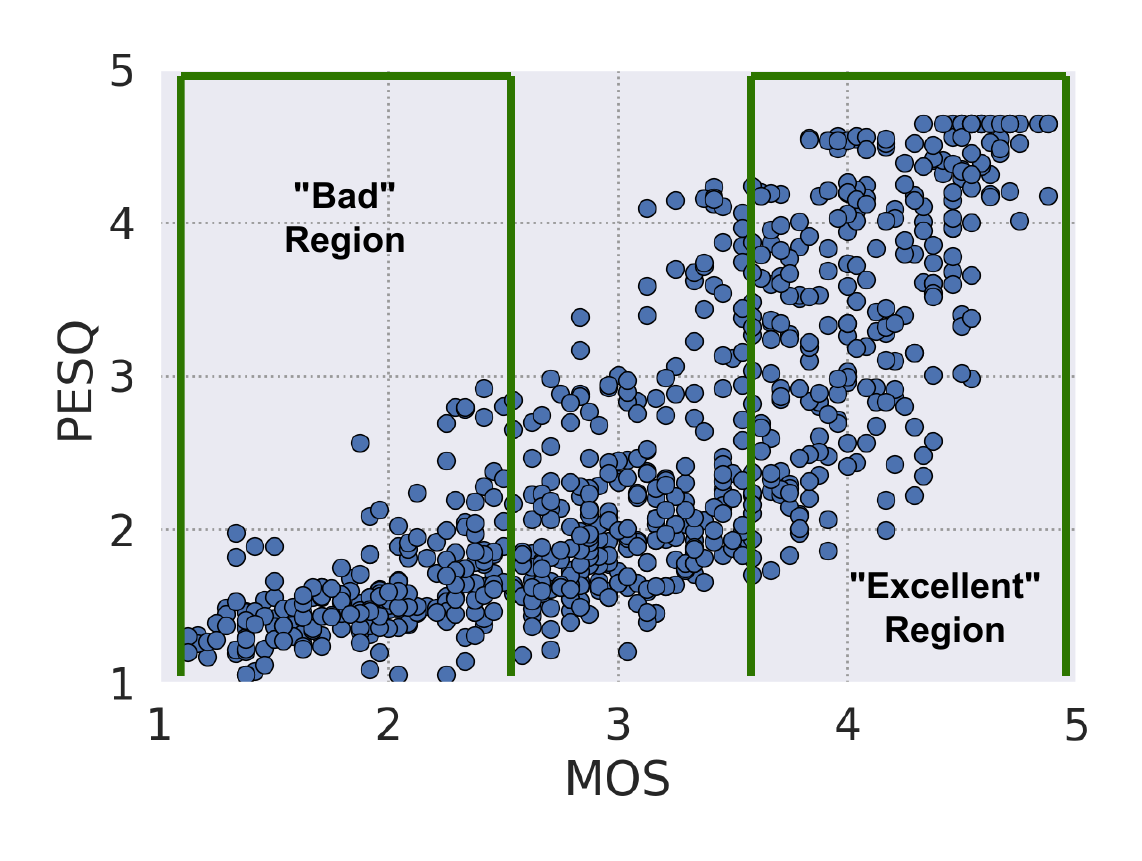}
  \caption{PESQ predictions on P23 EXP1 and restricted range computed by splitting the ground truth MOS values into 4 regions. Bad and excellent regions correspond to groups with the lowest and the highest MOS respectively.}
  \label{fig:PESQ_rr}
\end{figure}

\begin{figure*}[!t]
	\centering
        \begin{subfigure}{0.44\textwidth}
            \centering
		\includegraphics[width=\linewidth]{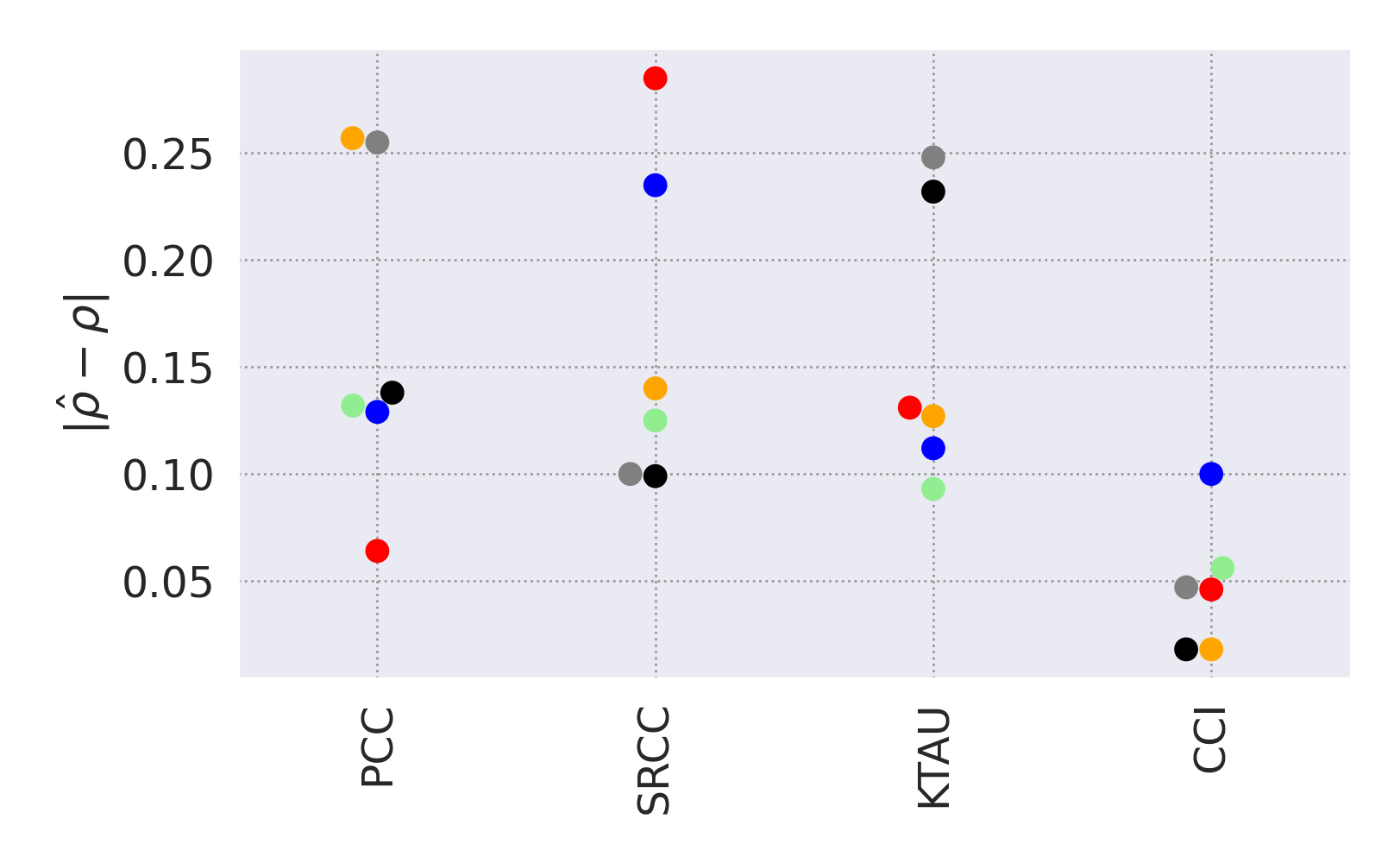}
		\caption{Restricted Range Bad, 2 Split}
		\label{fig:rr2bad}
	\end{subfigure}
	\hfill
        \begin{subfigure}{0.44\textwidth}
		\includegraphics[width=\linewidth]{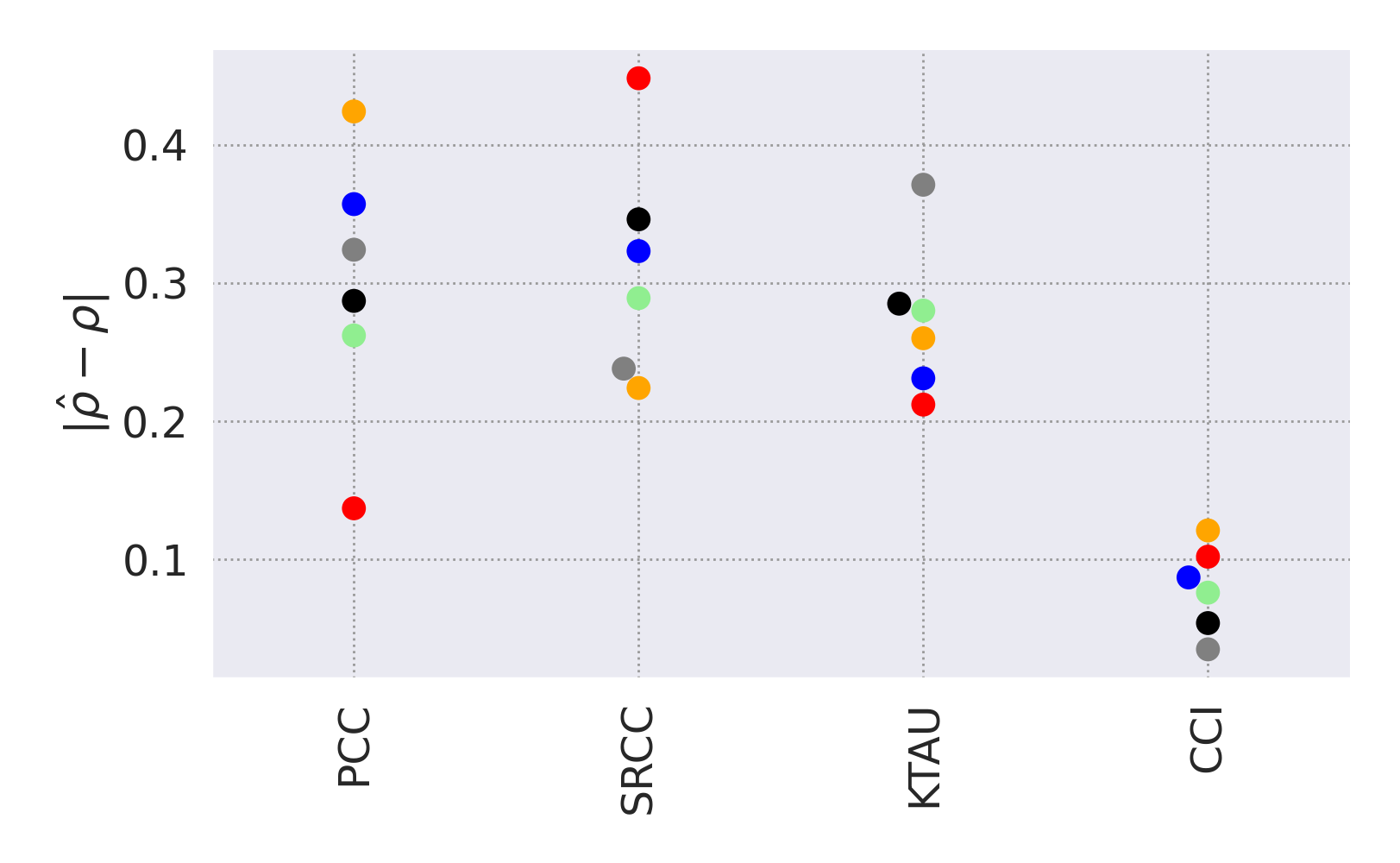}
		\caption{Restricted Range Excellent, 2 Split}
		\label{fig:rr2excellent}
	\end{subfigure}
        
        \vskip\baselineskip
        
	\begin{subfigure}{0.44\textwidth}
	        \includegraphics[width=\linewidth]{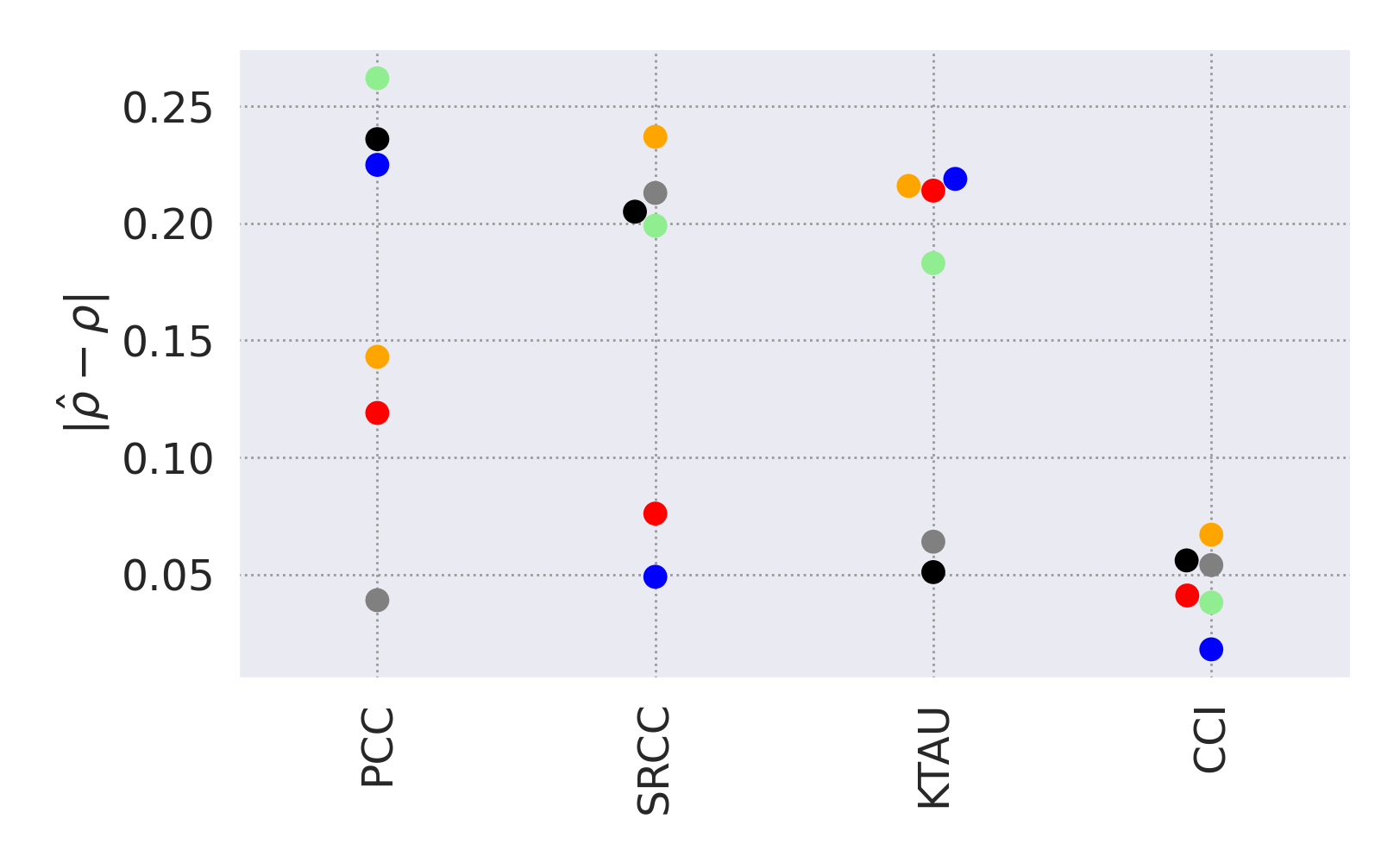}
	        \caption{Restricted Range Bad, 4 Split}
	        \label{fig:rr4bad}
         \end{subfigure}
         \hfill
         \begin{subfigure}{0.44\textwidth}
	        \includegraphics[width=\linewidth]{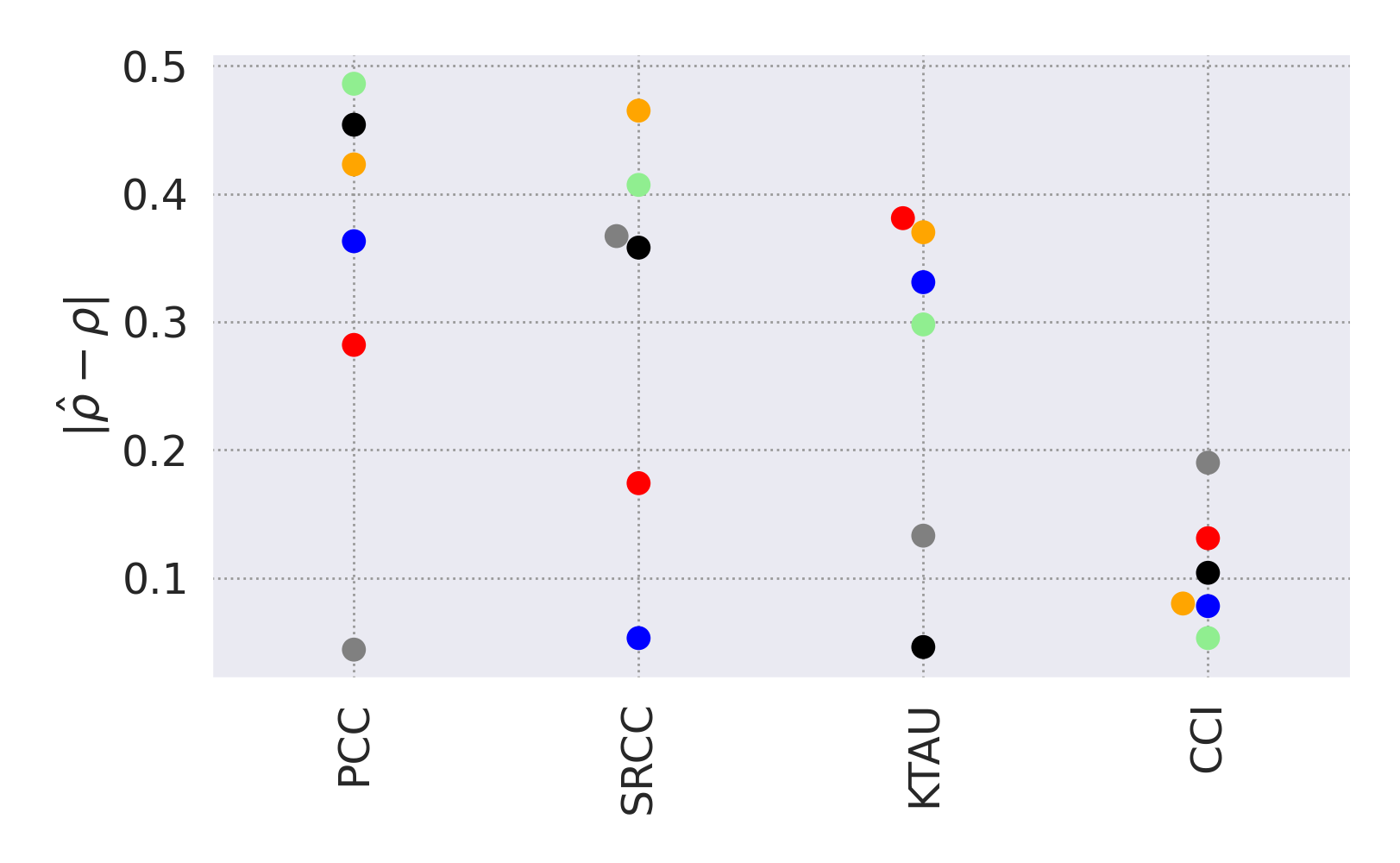}
	        \caption{Restricted Range Excellent, 4 split}
	        \label{fig:rr4excellent}
         \end{subfigure}
	\caption{Absolute difference between statistical metrics in a restricted range $\protect\widehat{\protect\rho}$ and the whole dataset $\rho$ with the following metric-dataset: VISQOL $\protect\rightarrow$ P23EXP1 \protect\tikzbullet{blue}{blue} P23EXP3 \protect\tikzbullet{orange}{orange} TCD-VOIP \protect\tikzbullet{red}{red}, PESQ $\protect\rightarrow$ P23EXP1 \protect\tikzbullet{green}{green} P23EXP 3 \protect\tikzbullet{black}{black} TCD-VOIP \protect\tikzbullet{gray}{gray}}

\label{fig:restricted range}
\end{figure*}

In Figure~\ref{fig:exp1_pesq} we show statistics for the model PESQ using the database P23 EXP1.
Because we have 2 quality models, 3 databases, and 3 statistics, to avoid plotting 18 figures we aggregate them based on the sample size index which is a mapping of integers to each corresponding sample size as shown in Table \ref{tab:samplesizemapping}.
For each sample size index, we aggregate by taking the mean of the quality model scores i.e., PESQ and ViSQOL. Although this averaging process across varying stimuli numbers and databases should not be done when evaluating quality models, here it is done solely to illustrate the trends. Our analysis shows that regardless of the metric or database used, statistical metrics exhibit consistent behaviour in relation to sample size, mirroring the trend observed with PESQ on the P23 EXP1 dataset. For detailed trends of each metric and database combination, refer to the repository for individual plots.

It is expected that the sample mean gets closer to the population mean as the sample size increases. However, it is unknown to which extent statistical metrics deviate from the population mean.
Figure~\ref{fig:exp1_sample_mean} shows the absolute difference between the sample mean $\hat{\rho}$ and the population mean $\rho$ of PESQ and ViSQOL aggregated over the 3 databases.
We observe that SRCC highly suffers when using lower sample sizes and should be used with caution. PCC and KTAU sample means are not far from the population mean as well as our proposed metric CCI. However, in both Figure~\ref{fig:exp1_percepoint} and \ref{fig:exp1_std} we show how PCC and KTAU show higher standard deviation and extreme values (5th and 95th percentile) indicating less robustness against a particular subsample of stimuli. The proposed method CCI is the best and it shows very low dependence from a particular subset of stimuli given the lower standard deviation when sampling 1000 different subsamples. 
Our analysis shows that PCC is the best among the commonly used statistical metrics (PCC, SRCC, and KTAU) and that the proposed method CCI is further better. We observe how for some sample sizes the difference between PCC and CCI almost doubles e.g., CCI$\approx0.02$ and PCC$\approx0.04$ for sample size index 12.

\subsection{Experiment 2: Rater Sampling Variability}
In this experiment, we address the problem of evaluating the robustness of statistical metrics when different rater groups are sampled. We take 8 linearly spaced rater sizes from 12 to 20 and for each rater size, we draw 1000 different combinations of rater subsets. To avoid a selection bias, the combination of raters for the samples are taken with replacement with respect to the population size. For example, combination 1 of rater size 12 might include a certain rater A that it can be excluded in the combination 2. This approach makes sure that every rater is uniformly sampled from the population and that samples of raters are not biased towards particular raters.
In this experiment, we do not measure how statistical metrics are accurate compared to the population parameters since when taking fewer raters it is expected to have different MOS values. However, the standard deviation can still indicate the robustness of considering different rater groups. We measure the standard deviation over the 1000 drawn groups for each fixed rater size as shown in Figure~\ref{fig:exp2_std}. Results are aggregated with the average as done in Experiment~1. However, here we do not need to use index mapping since we have the same number of raters for each dataset used.

Results show that when decreasing the rater size, higher uncertainty can be observed in all the statistical metrics. Among the traditional statistical metrics, SRCC is slightly more robust than PCC and KTAU. It can be observed that KTAU, SRCC, and PCC show a higher spread than the proposed CCI. These findings reveal that a subset of rater can significantly deviate the performance of quality models. This can be problematic for crowdsourcing where all the stimuli cannot be assigned to all the raters and different subgroups evaluate different stimuli. The proposed CCI show a lower standard deviation indicating higher robustness against rater uncertainty. Our results indicate that adding or removing a few raters could mislead researchers' conclusions on quality models when using PCC, SRCC, and KTAU. The proposed CCI mitigates this issue showing higher robustness.

\subsection{Experiment 3: Restricted Range}
Restriction of range occurs when PCC and SRCC show significant deviation when calculated within a subset of data points. For example, Figure \ref{fig:PESQ_rr} shows PESQ scores computed on P23 EXP1 split into 4 different groups based on MOS percentile. We identify the first region that we call 'Bad' and the high-quality region that we call 'Excellent', so to follow the naming convention of the ACR scale (Bad, Poor, Fair, Good, Excellent). When computing statistical metrics on these subintervals, they significantly drop.
Consequently, assessing quality within a confined range may yield results that do not accurately represent the true performance. A lower correlation with MOS is observed because of the imprecise quality range in the collected labels. Since each quality range is specific to the dataset, it is uncertain beforehand whether a particular quality database's range effectively captures the application being studied. 

\begin{figure*}[!t]
	\centering
	\begin{subfigure}{0.32\textwidth}
            \centering
		\includegraphics[width=\linewidth]{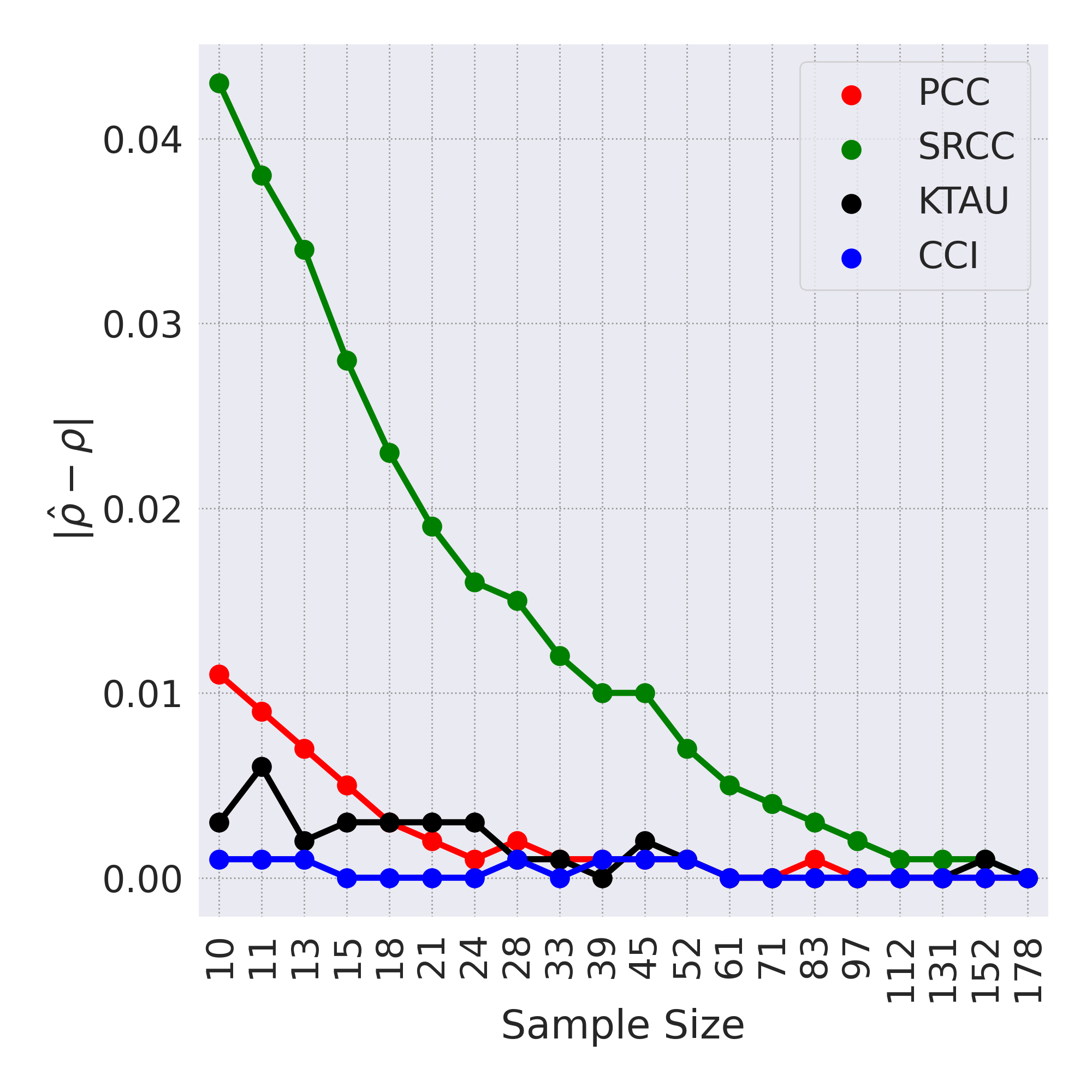}
		\caption{Sample Mean}
		\label{fig:exp1_ssim_sample_mean}
	\end{subfigure}
	\begin{subfigure}{0.32\textwidth}
		\includegraphics[width=\linewidth]{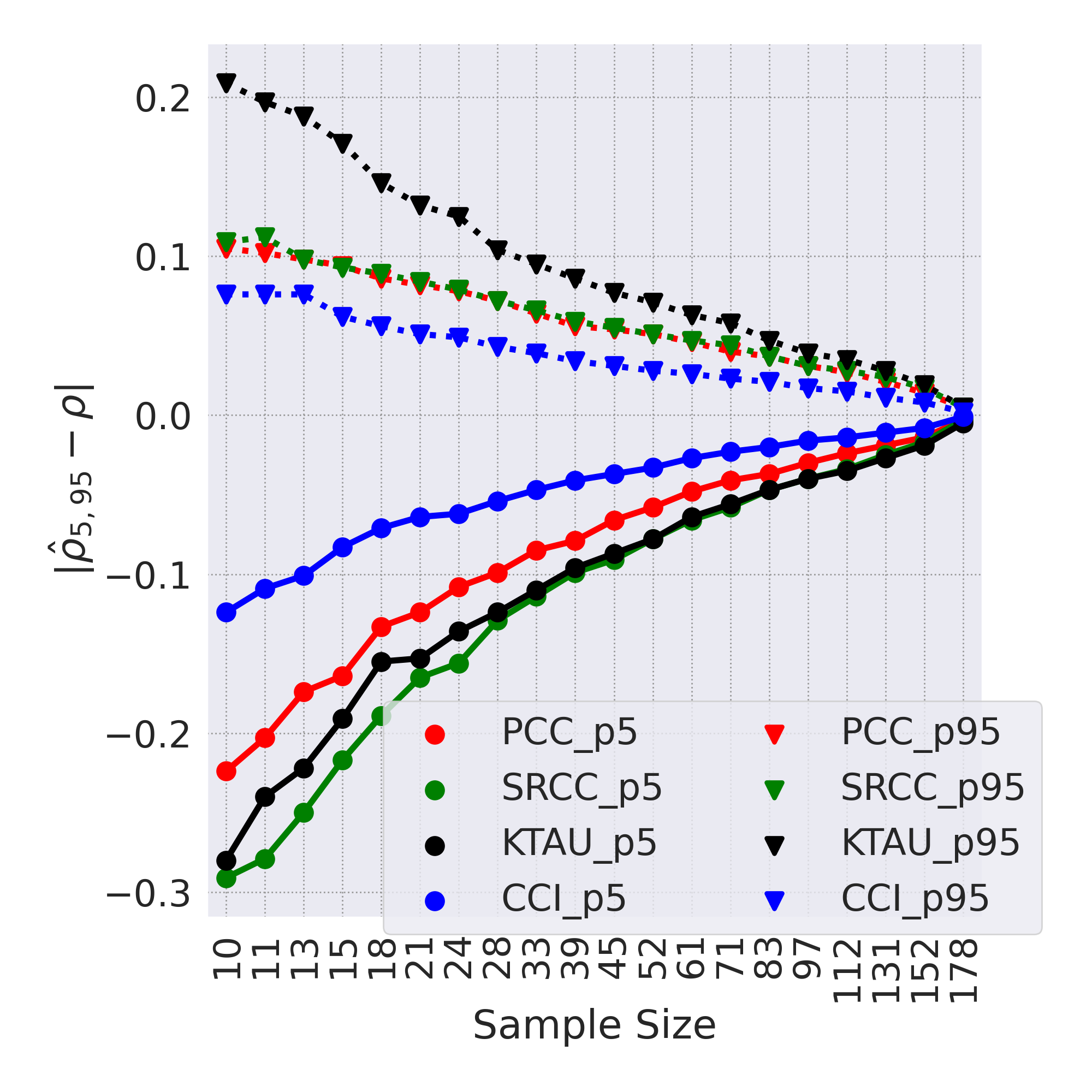}
		\caption{5th and 95th Percentile Deviation}
		\label{fig:exp1_ssim_percepoint}
	\end{subfigure}
	\begin{subfigure}{0.32\textwidth}
	        \includegraphics[width=\linewidth]{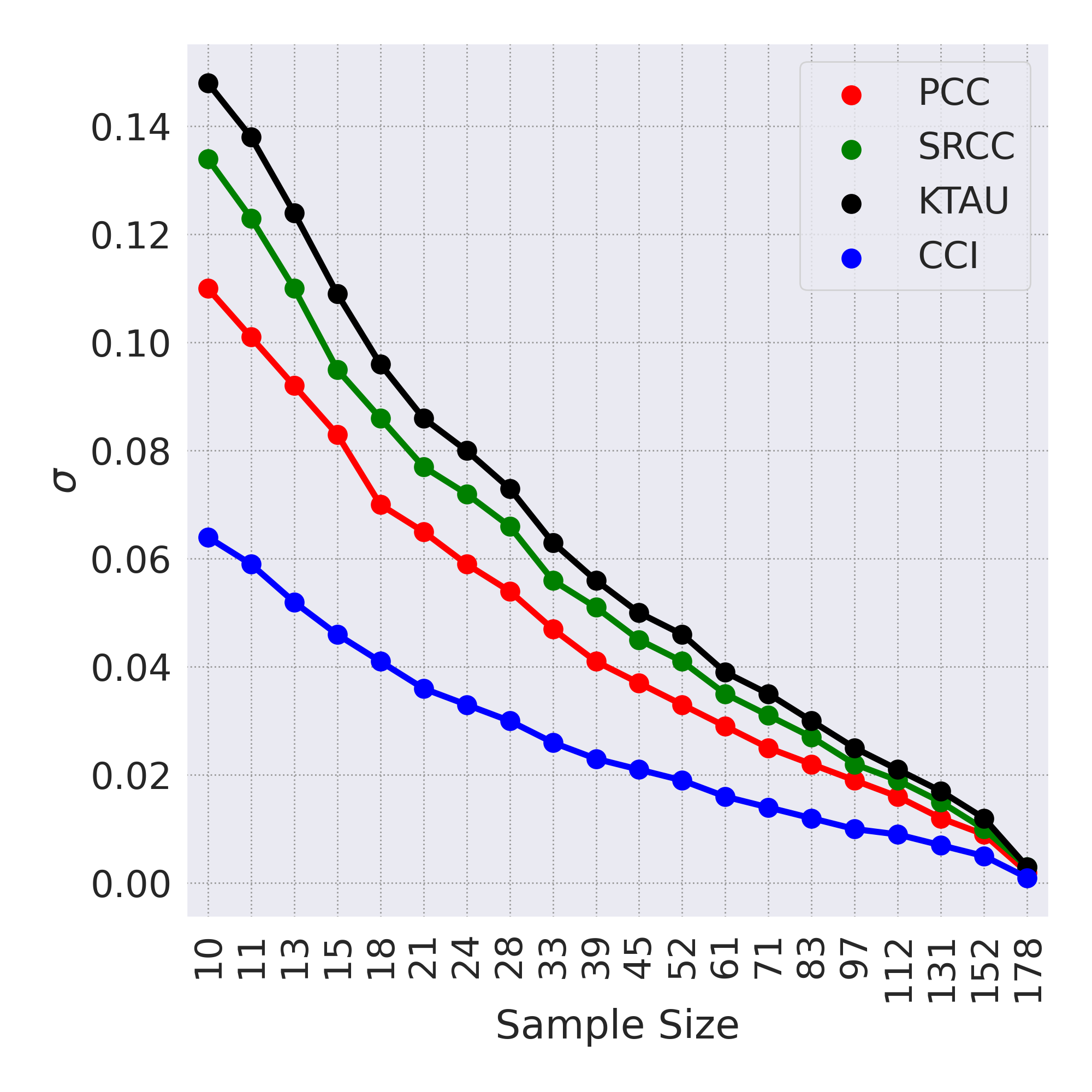}
	        \caption{Standard Deviation}
	        \label{fig:exp1_ssim_std}
         \end{subfigure}
	\caption{(a) Sample mean, (b) 5th and 95th percentiles, and (c) standard deviation, using SSIM calculated on JPEG XR.} 
	\label{fig:exp1_aggregated_ssim}
\end{figure*}

\begin{figure}[!t]
  \centering
  \includegraphics[width=0.80\linewidth]{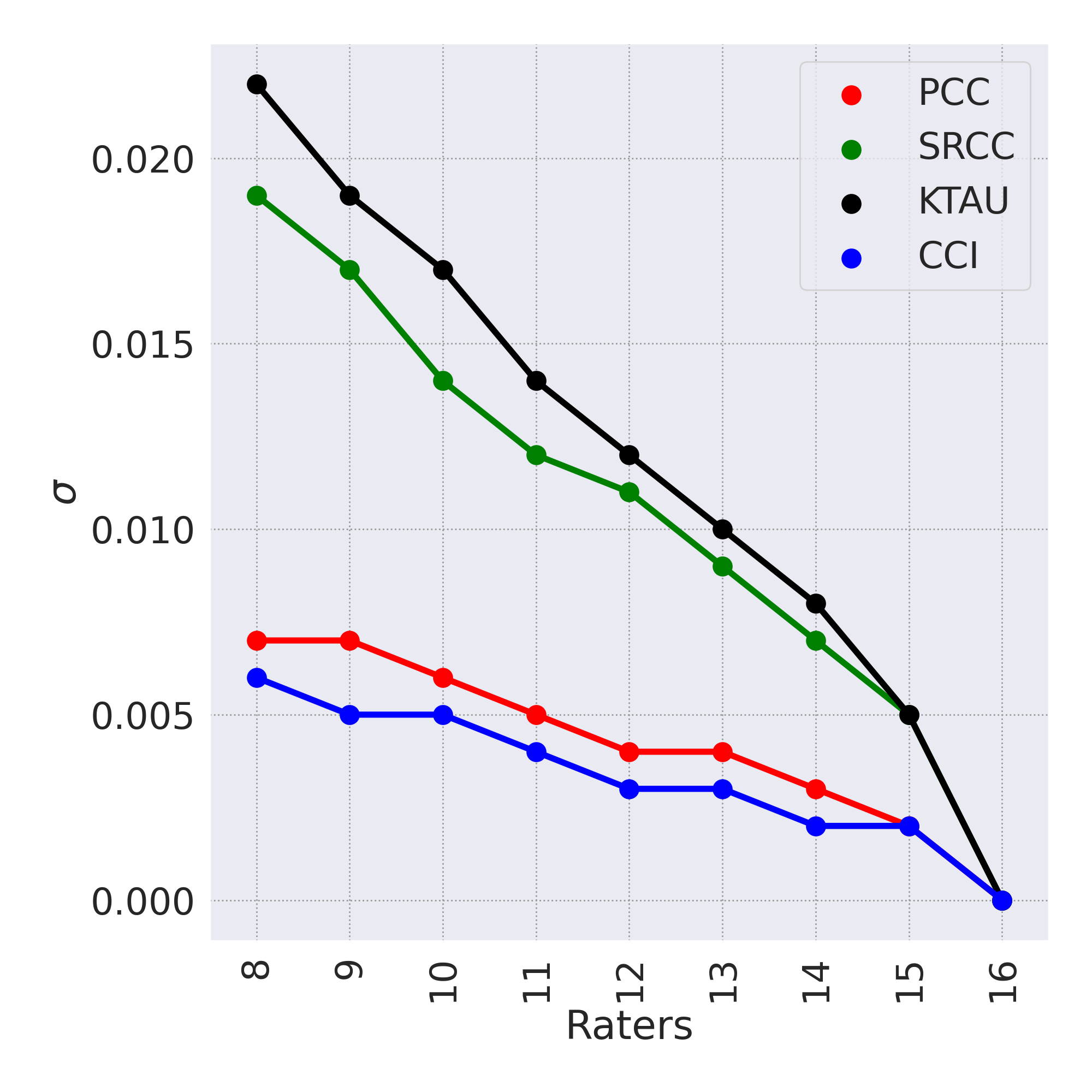}
  \caption{Standard deviation computed on 1000 subrater groups for each rater size. Statistical metrics calculated using SSIM predictions and MOS scores of the image dataset JPEG XR}
  \label{fig:exp2_std_SSIM}
\end{figure}

To examine how statistical metrics perform under range restriction, we adopt two methodologies for segmenting the MOS: a 2-split and a 4-split approach. The 2-split method divides the data at the median, labelling all points below as 'Bad' and those above as 'Excellent'. Meanwhile, the 4-split method partitions the MOS into four groups, where the lowest is labelled as 'Bad' and highest as 'Excellent' quality segments for analysis, as demonstrated in Figure~\ref{fig:PESQ_rr}. Figure~\ref{fig:restricted range} compares the statistical metrics derived from these subregions, $\widehat{\rho}$, against those calculated over the entire dataset, $\rho$. Our findings reveal that traditional correlation measures significantly deviate under range restriction, highlighting their vulnerability. However, our proposed CCI  demonstrates robustness in these conditions, maintaining stability across varied dataset groups.
Our analysis suggests that using CCI can be convenient for applications where the quality range might be uncertain as in the wild conditions where we do not know in advance which type of degradations are encountered. In these scenarios, there is a high chance to have an uncalibrated quality range which will lead to restriction of range and wrong conclusions.

\subsection{Experiment 4: Image Quality Assessment Evaluation}
\begin{figure}[!t]
	\centering
        \begin{subfigure}{0.24\textwidth}
		\includegraphics[width=\linewidth]{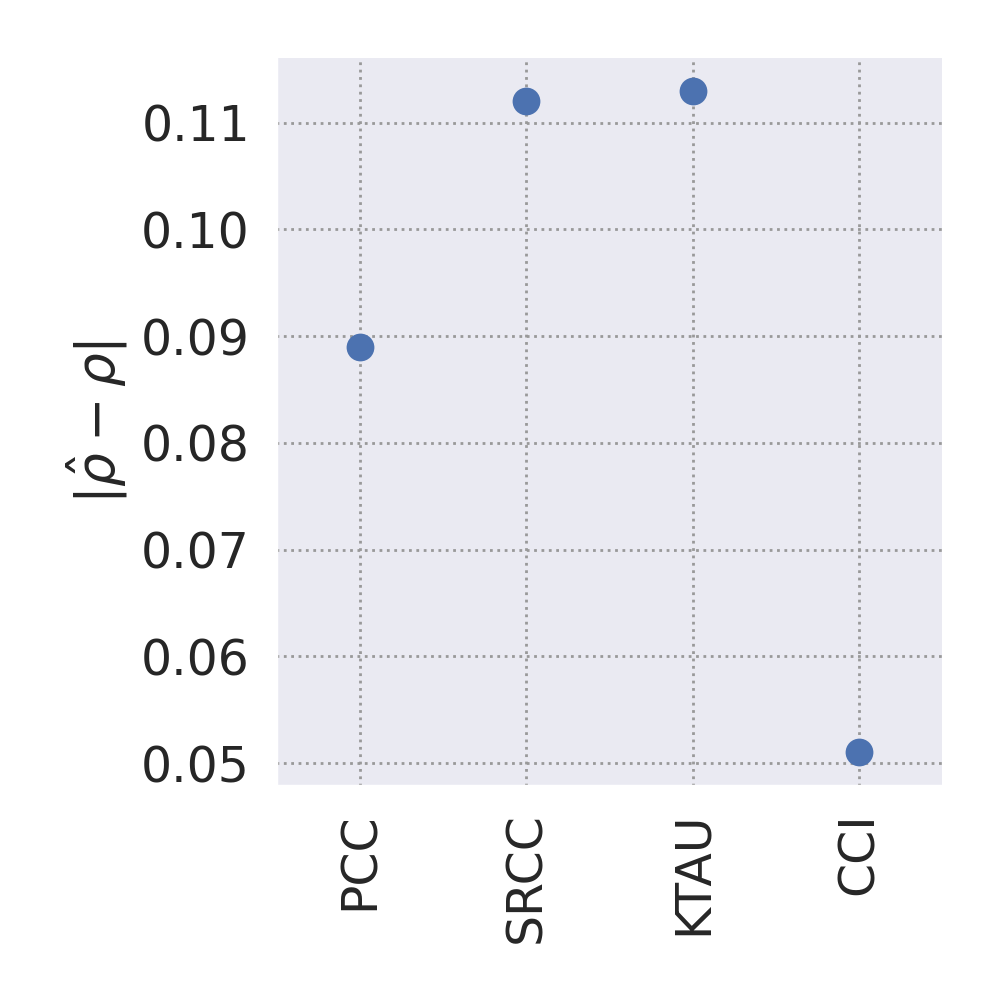}
		\caption{Bad, 2 Split}
		\label{fig:rr2badSSIM}
	\end{subfigure}
        \begin{subfigure}{0.24\textwidth}
		\includegraphics[width=\linewidth]{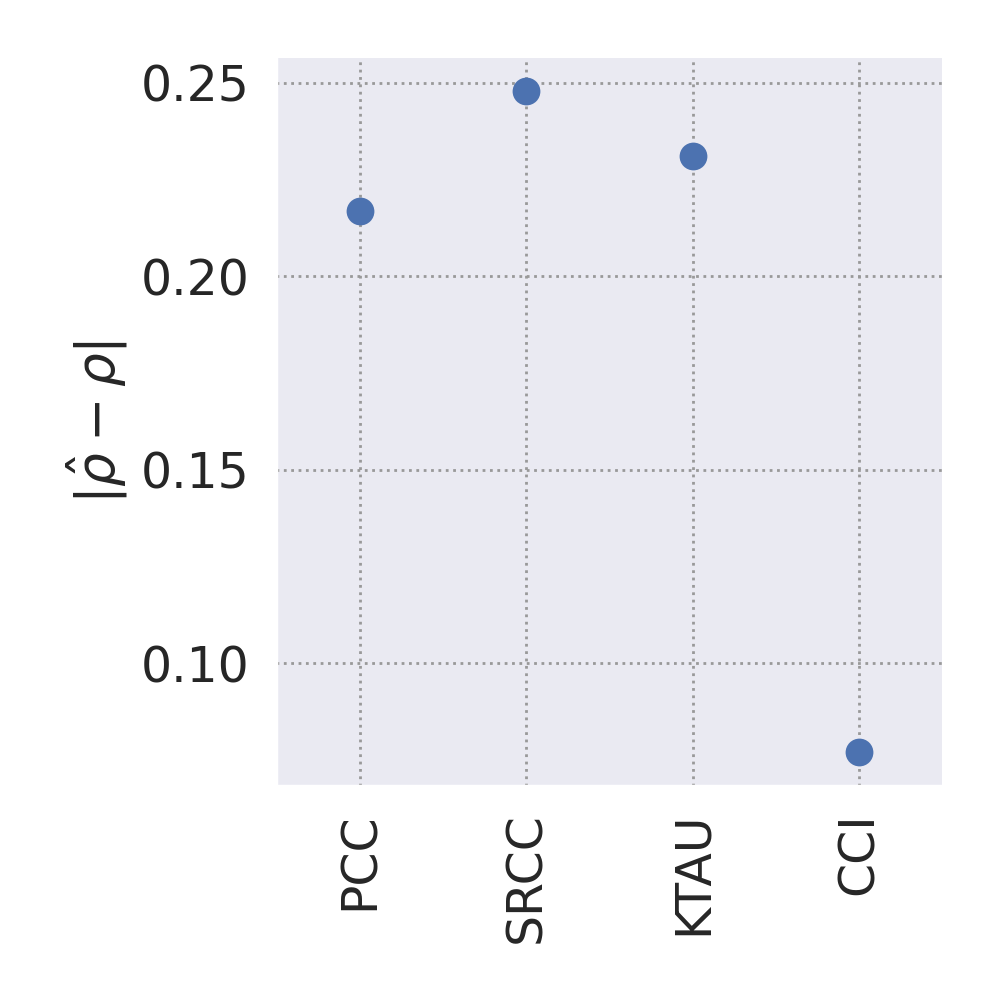}
		\caption{Excellent, 2 Split}
		\label{fig:rr2excellentSSIM}
	\end{subfigure}
        
        
	\begin{subfigure}{0.24\textwidth}
	        \includegraphics[width=\linewidth]{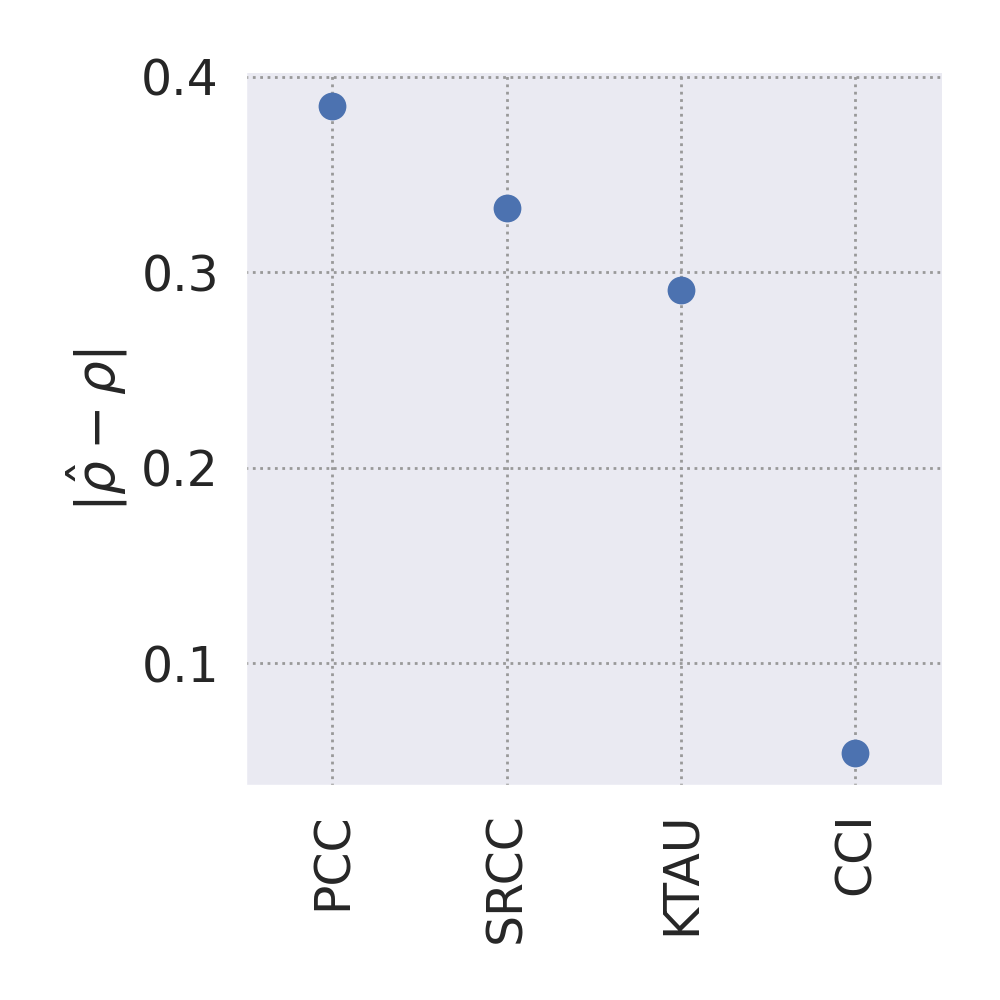}
	        \caption{Bad, 4 Split}
	        \label{fig:rr4badSSIM}
         \end{subfigure}
         \begin{subfigure}{0.24\textwidth}
	        \includegraphics[width=\linewidth]{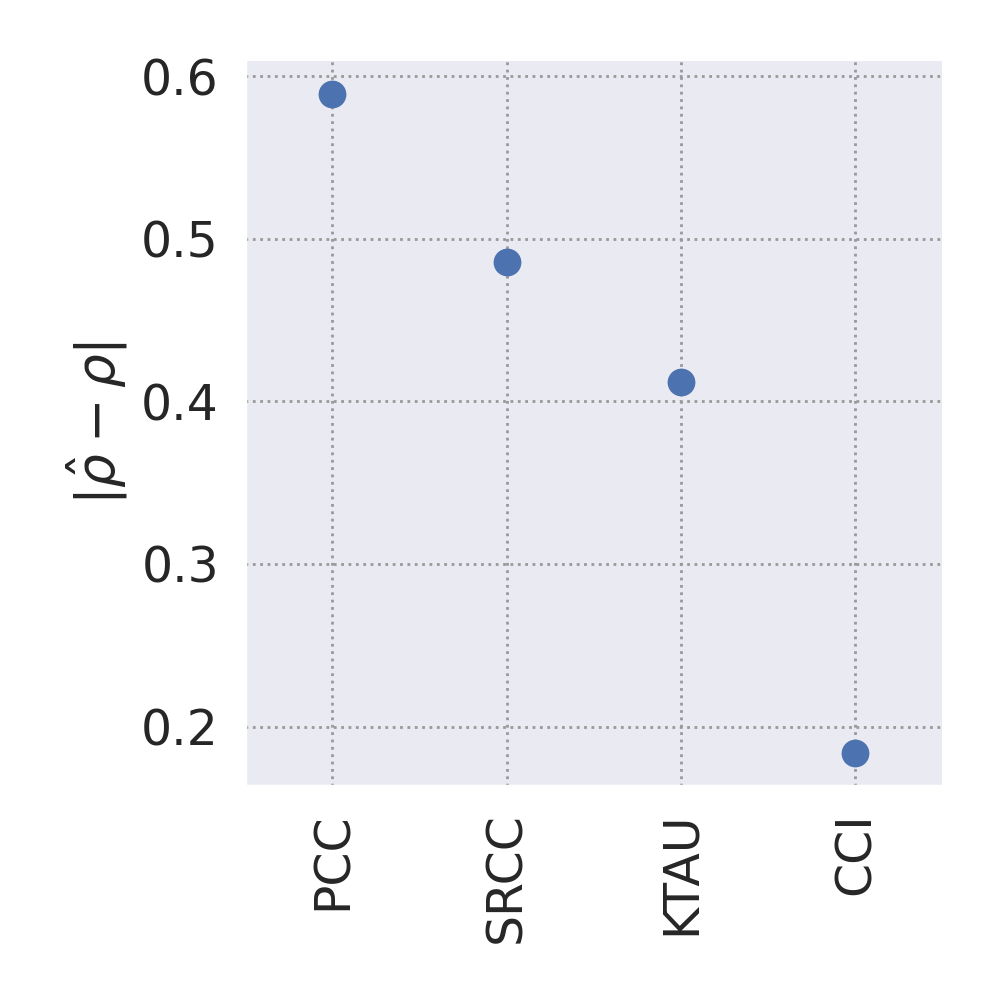}
	        \caption{Excellent, 4 Split}
	        \label{fig:rr4excellentSSIM}
         \end{subfigure}
	\caption{Absolute difference between statistical metrics in a restricted range $\widehat{\rho}$ and the whole dataset $\rho$ using SSIM computed on the JPEG XR dataset}

\label{fig:restricted range ssim}
\end{figure}

In this section, we replicate the three experiments done with speech using the SSIM, which is a full-reference image quality model. We use the image dataset JPEG XR, which includes 186 stimuli of high-resolution JPEG images~\cite{de2009subjective}. Image ratings have been collected using a continuous scale from 0 (Bad) to 100 (Excellent) to measure the perceived quality of 16 raters. Therefore, this experiment will also validate a perceptual scale with a higher resolution compared to the ACR scale used to rate the data used for the speech experiments. We exclude 6 stimuli representing quality scores calculated on the reference images.
The three experiments are shown in Figure~\ref{fig:exp1_aggregated_ssim}, Figure~\ref{fig:exp2_std_SSIM}, and Figure~\ref{fig:restricted range ssim} respectively.

We observe that statistical metrics follow the same trend as the speech data. This means that the pitfalls of statistical metrics are also present when perceptual scales with a higher resolution than ACR are used. In addition, the SSIM experiments suggest that the media used does not affect the observed weakness of statistical metrics discussed in this paper. These results imply that the proposed CCI can be utilized for more multimedia quality assessment tasks e.g., virtual reality and video QoE. 

\begin{table*}[!t]
\caption[C]{Statistical metric population parameters of PESQ and ViSQOL computed per file.}
\centering
\Huge
\ra{1.1}
\begin{adjustbox}{max width=0.80\textwidth}
\fontsize{24pt}{24pt}\selectfont
\begin{tabular}{@{}lccccccccccccccccccccccccccccccccccccccccccccccccccc@{}}\toprule

& \multicolumn{2}{c} {\phantom} & \multicolumn{5}{c} {\textbf{P23 EXP 1}} & \multicolumn{8}{c}{\textbf{P23 EXP 3}} & \multicolumn{8}{c}{\textbf{TCD-VoIP}}\\
\cmidrule{2-8} \cmidrule{10-16} \cmidrule{18-24}

\textbf{Model} &  PCC && SRCC && KTAU && CCI && PCC && SRCC && KTAU && CCI && PCC && SRCC && KTAU && CCI & \\ \midrule
PESQ       & 0.84 && 0.90  && 0.73 && 0.96 && 0.81 && 0.79 && 0.61 && 0.93 && 0.90 && 0.90 && 0.72  && 0.95 \\
ViSQOL     & 0.82 && 0.82  && 0.63 && 0.91  && 0.75 && 0.72 && 0.56 && 0.87 && 0.82 && 0.82 && 0.63  && 0.90  \\
\bottomrule
\label{tab:MOS_pred}
\end{tabular}
\end{adjustbox}
\end{table*}

\section{CCI Interpretation}
In this section, we discuss how CCI should be interpreted and that, unlike traditional statistical metrics, it is easier to understand. In Table \ref{tab:MOS_pred} we show the file-based population parameters of PESQ and ViSQOL on the 3 databases P23 EXP1, P23 EXP3, and TCD-VOIP. 
We observe that the proposed CCI is aligned with the other statistical metrics, indicating that PESQ is superior to ViSQOL for these 3 databases. Unlike other metrics, the CCI is easy to interpret. For example, ViSQOL for TCD-VOIP shows CCI=90, which means that 90\% of statistically different pairs are correctly ranked while the remaining 10\% is where the quality model fails. Other metrics are more difficult to interpret. Our previous simulated analysis in Section \ref{sec:experiments} has shown that CCI is more robust than other statistical metrics. This means that shifting from CCI=0.90 to CCI=0.92 can be meaningful unlike traditional statistical metrics (PCC, SRCC, and KTAU) where it could be due to chance.

\begin{figure}[!t]
  \centering
  \includegraphics[width=0.95\linewidth]{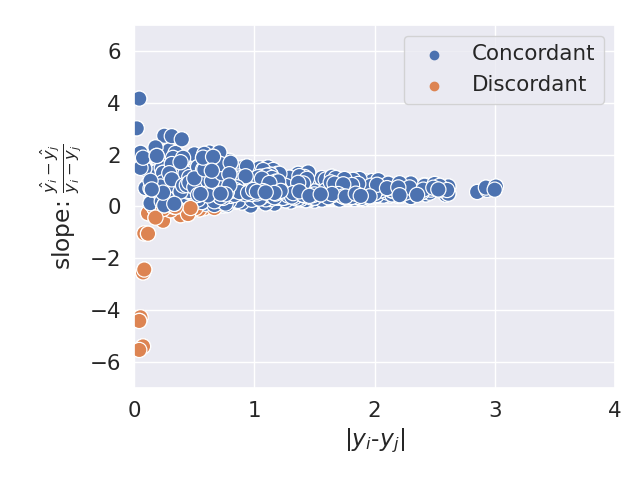}
  \caption{CCI visualisation: Per-condition PESQ, P23 EXP1}
  \label{fig:cci_vis_p23}
\end{figure}

\begin{figure}[!t]
  \centering
  \includegraphics[width=0.95\linewidth]{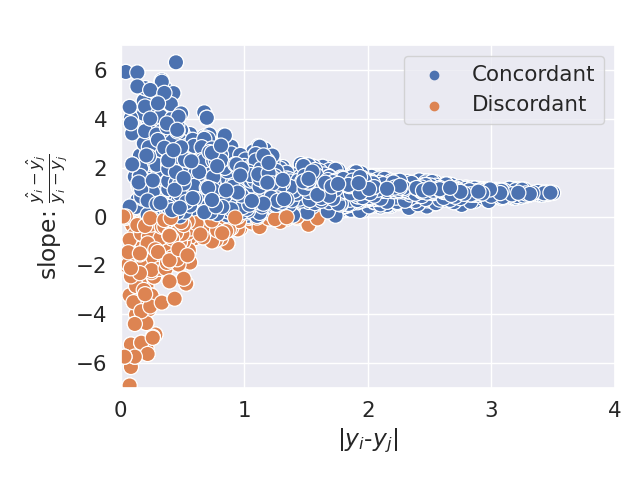}
  \caption{CCI visualisation: Per-condition PESQ, TCD VOIP}
  \label{fig:cci_vis_tcd}
\end{figure}

Typically PCC can be visualised using a scatter plot where predicted and ground truth MOS determine the coordinates of each point. To have insight into CCI, we propose to plot the concordant vs discordant in the constrained set by using the slope on the y-axis and the ground truth MOS distance on the x-axis. We plot PESQ per-condition on P23 EXP1 and TCD VOIP in Figure \ref{fig:cci_vis_p23} and Figure \ref{fig:cci_vis_tcd} respectively. 
We introduced the slope plot also in our previous paper~\cite{ragano2023audio} using the old CCI version.
It is possible to observe how PESQ works significantly better on P23 EXP1 when evaluated per condition. Also, it is clear how the majority of discordant pairs for TCD VOIP occur for very close MOS where quality is more difficult to predict. These plots are generated using the constrained set only i.e., the one where pair confidence intervals do not overlap. The CCI shows how even when using statistically different MOS pairs, quality models might fail. We believe that these are the pairs that should have importance in the evaluation, where MOS uncertainty is little.

\section{Discussion}
\label{sec:discussion}
The key findings of this study reveal issues of typical statistical metrics (PCC, SRCC, and KTAU) for evaluating instrumental quality models when reducing sample size, varying raters, and restriction of range. Our analysis consistently demonstrated that these coefficients might fail to capture performances if these conditions are not checked. This finding aligns with some previous analyses of PCC and SRCC, shedding light on the underlying dynamics in the context of instrumental quality models. To solve these issues, we proposed the Constrained Concordance Index (CCI) which is based on incorporating rater subjectivity in the evaluation. We demonstrated that the proposed CCI does not suffer from the same issues of PCC, SRCC, and KTAU on instrumental quality models.

In comparing our results to previous studies, our findings extend existing literature in the domain of QoE and how to better quantify human perception subjectivity~\cite{seufert2019fundamental,hossfeld2011sos,chinen2021marginal,krasula2016accuracy}. While Krasula et al.~\cite{krasula2016accuracy} also incorporated rater subjectivity, our study aims to address different problems and scenarios namely: low sample sizes, rater variability, and restriction of range. 

The rise of new media technologies (e.g., VR, plenoptic) goes beyond traditional perception experience, which might challenge traditional quality assessment tools. Recently, the JPEG Pleno committee launched an initiative to implement new quality assessment standards~\cite{jpegpleno2022}, which seeks to explore new assessment protocols and tools that better adapt the emerging media. In this context, the proposed CCI could be an appealing alternative in scenarios where traditional statistical tools are not capturing performance differences of assessment models due to inherent differences during subjective experiments (e.g., quality range, group variability).

CCI is a useful statistical metric that can aid the development of new datasets with unconventional methods e.g., the ones that do not follow a lab-based protocol. Among the examples, we suggest crowdsourcing, the fusion of multiple datasets, heterogeneous subjects or labels of questionable reliability, when dataset conditions are collected in the wild and not carefully curated through lab-based pilot studies. 
In addition, CCI is useful to avoid potential high performance due to deep learning overfitting. In this scenario, a quality model could simply overfit the MOS uncertainty and show very good performance using traditional statistical metrics. With CCI, objective models trained on out-of-domain data can be evaluated in a fairer way since MOS uncertainty is taken into account.

\section{Conclusions}
\label{sec:conclusions}
This study tested the hypothesis that typical statistical metrics cannot be fully relied upon in some scenarios. Approaches to dataset labelling, such as crowdsourcing are becoming more common, where uncertainty in the ground truth and the predicted target need to be understood together. We explored the underlying latent factors behind human labelling in the context of QoE applications. We demonstrated pitfalls in conventional statistical metrics for the following scenarios: low sample size, restriction of range, and rater group sampling. We propose a new statistical metric, CCI, that addresses these issues through leveraging stimuli pairs where quality difference has high confidence. We believe that CCI can to be extended to other domains beyond speech and image such as immersive media QoE where a higher number of participants is required to collect reliable quality labels. It may also be useful in domains beyond quality assessment.
\bibliographystyle{IEEEtran}
\bibliography{sample-base2}
\end{document}